\documentclass[a4paper,fleqn,usenatbib]{mnras}

\usepackage{newtxtext,newtxmath}
\usepackage[T1]{fontenc}
\usepackage{ae,aecompl}

\usepackage{graphicx}	
\usepackage{amsmath}	
\usepackage{cleveref}

\usepackage{subfig}
\usepackage[utf8]{inputenc}		
\usepackage{pifont}
\usepackage{romannum}

\title{The onset of bar formation in a massive galaxy at $z \sim 3.8$}

\author[A. Amvrosiadis et al.]{A. Amvrosiadis$^{1}$\thanks{E-mail: aristeidis.amvrosiadis@durham.ac.uk},
S. Lange$^{1}$,
J. Nightingale$^{2, 1}$,
Q. He$^{1}$,
C. S. Frenk$^{1}$, 
K. A. Oman$^{1}$, 
I. Smail$^{2}$, \newauthor 
A. M. Swinbank$^{2}$, 
F. Fragkoudi$^{1}$,
D. A. Gadotti$^{2}$,
S. Cole$^{1}$,
E. Borsato$^{3}$, 
A. Robertson$^{4}$, 
R. Massey$^{1, 2}$, \newauthor
X. Cao$^{5, 6}$, 
R. Li$^{6, 5}$
\vspace{4mm}\\
$^{1}$ Institute for Computational Cosmology, Department of Physics, Durham University, South Road, Durham DH1 3LE, UK \\
$^{2}$ Centre for Extragalactic Astronomy, Department of Physics, Durham University, South Road, Durham DH1 3LE, UK \\
$^{3}$ Dipartimento di Fisica e Astronomia ``G. Galilei'', Università di Padova, vicolo dell’Osservatorio 3, I-35122 Padova, Italy \\
$^{4}$ Jet Propulsion Laboratory, California Institute of Technology, 4800 Oak Grove Drive, Pasadena, CA 91109, USA \\
$^{5}$ School of Astronomy and Space Science, University of Chinese Academy of Sciences, Beijing 100049, China \\
$^{6}$ National Astronomical Observatories, Chinese Academy of Sciences, 20A Datun Road, Chaoyang District, Beijing 100012, China
}

\pubyear{???}

\begin{document}
\pagenumbering{arabic}
\label{firstpage}
\maketitle

\begin{abstract}
  We examine the morphological and kinematical properties of SPT\nobreakdash-2147, a strongly lensed, massive, dusty, star-forming galaxy at $z = 3.762$. Combining data from JWST, HST, and ALMA, we study the galaxy's stellar emission, dust continuum and gas properties. The imaging reveals a central bar structure in the stars and gas embedded within an extended disc with a spiral arm-like feature. The kinematics confirm the presence of the bar and of the regularly rotating disc. Dynamical
  modeling yields a dynamical mass, ${M}_{\rm dyn} = (9.7 \pm 2.0) \times 10^{10}$~${\rm M}_{\odot}$, and a maximum rotational velocity to velocity dispersion ratio, $V / \sigma = 9.8 \pm 1.2$. From multi-band imaging we infer, via SED fitting, a stellar mass, ${M}_{\star} = (6.3 \pm 0.9) \times 10^{10}$~$\rm{M}_{\odot}$, and a star formation rate, ${\rm SFR} = 781 \pm 99$~${\rm M_{\odot} yr^{-1}}$, after correcting for magnification. Combining these measurements with the molecular gas mass, we derive a baryonic-to-total mass ratio of ${M}_{\rm bar} / {M}_{\rm dyn} = 0.9 \pm 0.2$ within 4.0 kpc. {This finding suggests that the formation of bars in galaxies begins earlier in the history of the Universe than previously thought and can also occur in galaxies with elevated gas fractions.}
\end{abstract}


\section{Introduction} \label{sec:section_1}

Amongst high redshift ($z >$1) galaxy populations, submillimeter
(sub-mm) galaxies \citep[SMGs;][]{1997ApJ...490L...5S,
  1998Natur.394..248B, 1998Natur.394..241H, 1999ApJ...515..518E}, or,
more generally, dusty star-forming galaxies
\citep[DSFGs;][]{2014PhR...541...45C} make up the high-mass tail of
the stellar mass function \citep{2020MNRAS.494.3828D}; using standard
assumptions, their masses are inferred to lie in the range,
${M}_{\star} \sim 10^{10} - 10^{11}$~M$_{\odot}$
\citep[e.g.][]{2015ApJ...806..110D, 2017A&A...597A...5M}. Despite
their large masses, they are still actively forming stars; their
inferred star formation rates are amongst the highest known,
${\rm SFR} \sim 10^2 - 10^3$~$\rm M_{\odot} yr^{-1}$
\citep[e.g.][]{2014MNRAS.438.1267S}\footnote{The masses and star
  formation rates quoted in this paper are derived assuming a standard
  initial mass function (IMF) from \cite{2003ApJ...586L.133C}. However,
  \cite{2005MNRAS.356.1191B} have argued that the observed galaxy
  number counts at 850 \micron\ can only be explained assuming a
  top-heavy initial mass function for the stars formed in bursts. In
  this case, the resulting stellar masses and star formation rates are
  significantly smaller than the values derived assuming a standard
  IMF.}. These extreme star formation rates are fuelled by the large
amounts of molecular gas present in these systems,
$M_{\rm gas} \sim $\,10$^{10}$--10$^{11}$~M$_{\odot}$
\citep[e.g.][]{2013MNRAS.429.3047B, 2021MNRAS.501.3926B}. If these
galaxies can sustain their star formation rates, they will consume all
of their available gas in $t_{\rm depl} < 1$ Gyr
\citep[][]{2013MNRAS.429.3047B, 2021MNRAS.501.3926B}, in many cases
almost doubling their current stellar masses given their gas mass fractions, $f_{\rm gas}$ = 20 --
50~per~cent \citep[][]{2013MNRAS.429.3047B, 2021MNRAS.501.3926B}. These properties alone suggest that these galaxies are likely to evolve into the most massive early-type galaxies \citep[ETGs; e.g.][Amvrosiadis et al. in prep]{2014ApJ...788..125S, 2021MNRAS.501.3926B}. Indeed, physically-motivated models of the formation and evolution of submillimeter galaxies in the $\Lambda$CDM cosmology explicitly show that bright submillimeter galaxies at high redshift are the precursors of the brightest elliptical galaxies today \citep{2005MNRAS.356.1191B}.

Identifying the nature of SMGs/DSFGs is challenging because of the
presence of substantial amounts of dust, which obscures their
starlight.  However, over the past decade, interferometric
observations at submillimeter/millimeter (sub-mm/mm) wavelengths, from ALMA and NOEMA, have greatly increased our understanding of these galaxies. The high resolution and high
sensitivity of these facilities has allowed us to study
the distribution of dust continuum emission, which serves as a direct
tracer of star formation since dust re-emits absorbed UV/optical
emission at longer wavelengths. Our findings suggest that these sources exhibit compact dust continuum morphologies,
with effective radii, $r_{\rm eff} \sim 1 - 2$ kpc, and have
distributions resembling those of disc galaxies, with an average
S\'{e}rsic index, $n \sim 1$ \citep[e.g.][]{2016ApJ...833..103H,
  2019MNRAS.490.4956G}. Pushing the capabilities of current
instruments to their limit, high signal-to-noise ratio (SNR) sub kpc-scale ($\sim 500$~${\rm pc}$) observations reveal, in some sources, structures consistent with those expected for bars or spiral arms
\citep[e.g.][]{2019ApJ...876..130H}.

Mapping the distribution of star formation sites in these galaxies
alone does not, however, give us a complete picture.  Fortunately,
besides observations of the dust continuum emission, spectroscopic data can be obtained with interferometers for various far-infrared
and sub-mm emission lines (e.g. CO, [C{\sc ii}]). These lines are
expected to be less affected by dust attenuation
\citep[e.g.][]{2018A&A...609A.130L} and trace gas in different
phases. One important finding regarding morphologies is that the
gas distributions are significantly more extended than the dust
continuum and display disc-like characteristics in most cases
\cite[e.g.][]{2012ApJ...760...11H}. This suggests that star formation
takes place in the central regions of these galaxies, which represent
a small part of a more extended disc.

In addition, emission line observations measure
the dynamical state of these galaxies, provided that they have sufficient resolution and sensitivity so as to minimize the
effect of beam smearing \citep[][]{2022A&A...667A...5R}. Such
observations have allowed us to identify regularly rotating discs in
the majority of cases that have been studied so far
\citep[e.g.][]{2014A&A...565A..59D, 2018ApJ...863...56C,
  2020Natur.584..201R, 2021Sci...372.1201T, 2021MNRAS.507.3952R, 2021A&A...647A.194F,
  2022A&A...667A...5R, 2023arXiv230316227R} but also, in a few cases, galaxies undergoing
interactions \citep[e.g.][]{2017ApJ...846..108C, 2020MNRAS.494.5542R}. These observations have important implications
for the mechanism(s) that causes the gas to funnel to the central starburst and so maintain the extreme star formation in these
massive galaxies. Indeed, if sufficient resolution, SNR and sample sizes can be obtained, we should be close to distinguishing between the case where gas collapses to the centre due to a merging event and the case where the radial inflows are triggered by a disc instability. From the observations so far it seems that both mechanisms are at play, as expected from galaxy formation models \citep[e.g.][]{2000MNRAS.319..168C}, although their relative contributions are still unconstrained due to poor statistics.

While the spatial distribution of star formation and the
distribution and kinematics of gas in SMGs are relatively
well established, our understanding of the distribution of the stellar
component is limited. The majority of studies attempting to
characterise stellar morphologies have, until recently, relied on
observations with the Hubble Space Telescope (HST). At $z > 1$, HST
observations probe rest-frame UV/optical wavelengths, which mainly
trace the younger stellar populations, and are severely affected by
the large amounts of dust present, giving us an obscured view of the
stellar morphologies. These studies find that the stellar light
distribution (rest-frame UV/optical) appears to be morphologically
very different from the dust continuum distribution. In general, the
stars appear to be significantly more extended than the gas and show
clumpy/irregular features \citep[e.g.][]{2010MNRAS.405..234S,
  2015ApJ...799..194C, 2019ApJ...879...54L}. It is important to note
that associating these irregularities exclusively to recent
interactions, as previous studies have often done, can result in
erroneous conclusions. A more comprehensive understanding of the
stellar component requires longer wavelength observations
(rest-frame near-infrared; NIR) that are less affected by dust
attenuation.

The launch of the James Webb Space Telescope (JWST) has opened up a
new window into the study of stellar morphologies in galaxies. JWST offers
high-resolution and high-sensitivity imaging and spectroscopy
capabilities in the rest-frame NIR, which is less
affected by the copious amounts of dust present, especially in SMGs/DSFGs. Already we are
beginning to observe an increased fraction of disc galaxies at high
redshift \citep[e.g.][]{2022ApJ...938L..24F, 2022ApJ...938L...2F,
  2023ApJ...948L..18N}, many of which have well-defined stellar
structures such as spiral arms and bars, that were previously
classified as irregular based on HST observations
\citep[e.g.][]{2022ApJ...939L...7C, 2023ApJ...945L..10G}. In addition,
multi-band imaging with both NIRCam and MIRI allows accurate
determinations of stellar masses through
modelling their spectral energy distributions (SED), when previously
these were very uncertain
\citep[e.g.][]{2014A&A...571A..75M}.

Previous studies of SMGs have highlighted their complex and diverse
properties and the importance of studying them across a wide range of
wavelengths. {Our study will contribute to this effort by providing
detailed information on the internal structure and kinematics of SPT\nobreakdash-2147, a $z = 3.762$, massive DSFG, strongly lensed by a, z = 0.84, early-type galaxy (ETG). SPT\nobreakdash-2147 was originally
discovered in the South Pole Telescope (SPT) survey
\citep[][]{2013Natur.495..344V} and subsequently followed-up with many facilities.} We will use a combination of data
from JWST, HST and ALMA and, with the aid of
strong gravitational lensing, study the different components
of the galaxy and its interstellar medium (ISM) down to sub-kpc scales.

The outline of the paper is as follows; In
Section~\ref{sec:section_2}, we introduce the various datasets used in
this work and derive the basic (galaxy-integrated) physical properties
of our source. In Section~\ref{sec:section_3}, we describe the details
of our strong lens modelling analysis. In
Section~\ref{sec:section_4}, we present a morphological and
kinematical analysis of our source, which we further discuss in
Section~\ref{sec:section_5}. Finally, in Section~\ref{sec:section_6}
we provide a summary of our findings. Throughout this work, we adopt a
spatially-flat $\Lambda$-CDM cosmology with H$_0=67.8 \pm 0.9\,\mathrm{km}\,\mathrm{s}^{-1}\,\mathrm{Mpc}^{-1}$ and $\Omega_{\rm M}=0.308 \pm 0.012$
\citep{2016A&A...594A..13P}.

\section{Observations \& source properties} \label{sec:section_2}

\begin{figure*}
    \centering
    \includegraphics[width=0.875\textwidth, height=0.685\textheight]{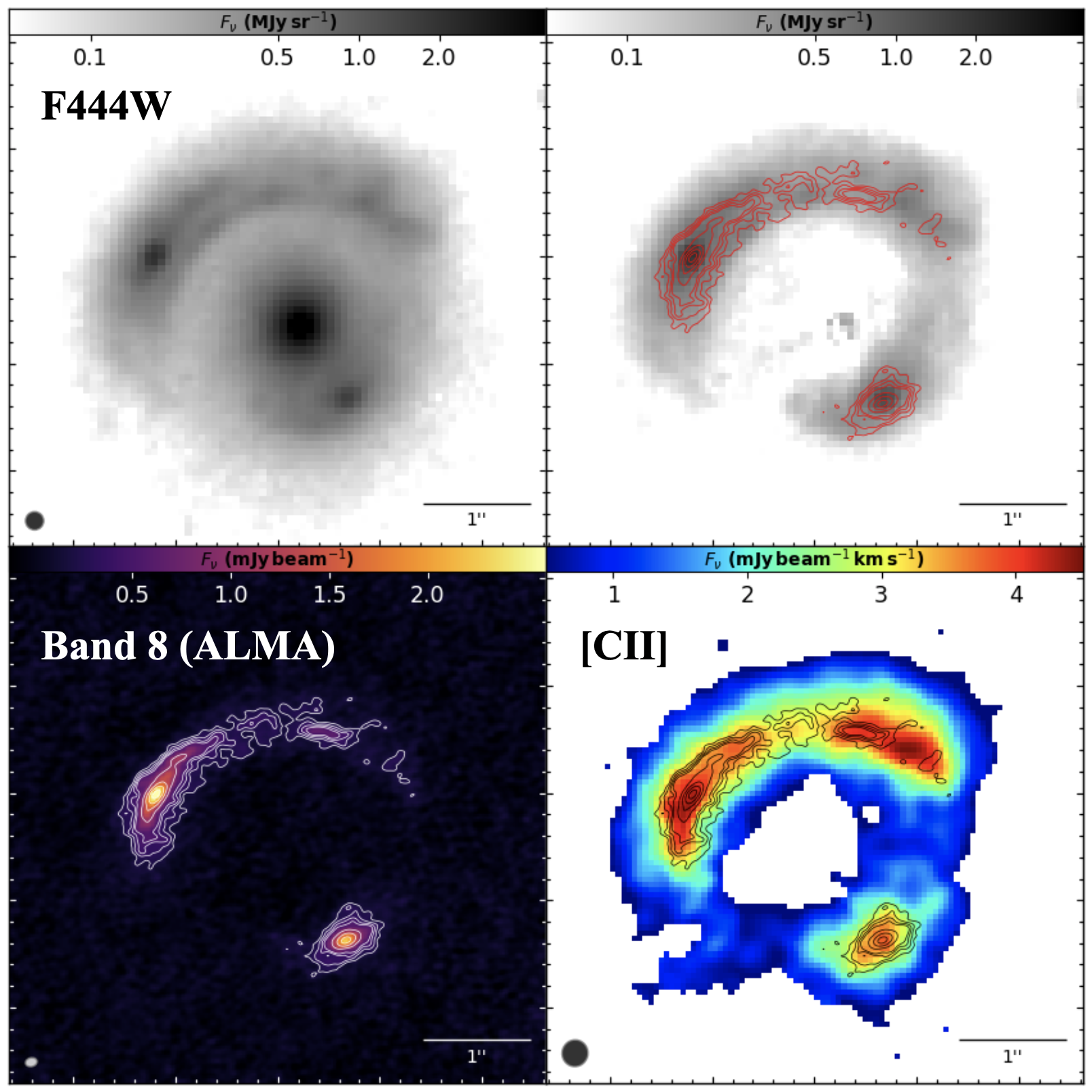}
    \caption{\textit{(Top left):} JWST/NIRCam stellar continuum images in the F444W filter ($\lambda_{\rm rest} \sim 0.92 \, {\rm \mu m}$). \textit{(Top right):} Same as the top left panel but with the light from the lens galaxy subtracted (see Section~\ref{sec:lens_light}). \textit{(Bottom left):} ALMA dust continuum images in band 8 ($\lambda_{\rm obs}\sim\,$742$\,\mu$m). \textit{(Bottom right):} [C{\sc ii}] velocity-integrated emission line image ($\lambda_{\rm rest} =\,$157.74$\,\mu$m), which was produced by fitting a Gaussian distribution to the spectrum in each individual pixel of the observed cube (see main text for details). The red, white and black contours in the top right, bottom left and bottom right panels, respectively, correspond to the dust continuum emission. The ellipses at the bottom left corners of the each panel show the size of the resolution of these observations; in the case of ALMA this corresponds to the synthesized beam. North is up and East is left in all panels.}
    \label{fig:fig1}
\end{figure*}

In this section, we provide an overview of the available datasets and derive fundamental galaxy-integrated physical properties for SPT\nobreakdash-2147.

\subsection{Data}

An mentioned above, we utilize data obtained from multiple facilities. These datasets allow us to examine various constituents of this galaxy: stars, dust, and gas. In Figure~\ref{fig:fig1}, we present a visual comparison of these distinct components side by side in the image plane. Only a selection of the available data, specifically those that are primarily being used later on in the analysis, are shown in this figure.

\subsubsection{JWST}

\begin{figure*}
    \centering
    \includegraphics[width=0.95\textwidth,height=0.19\textheight]{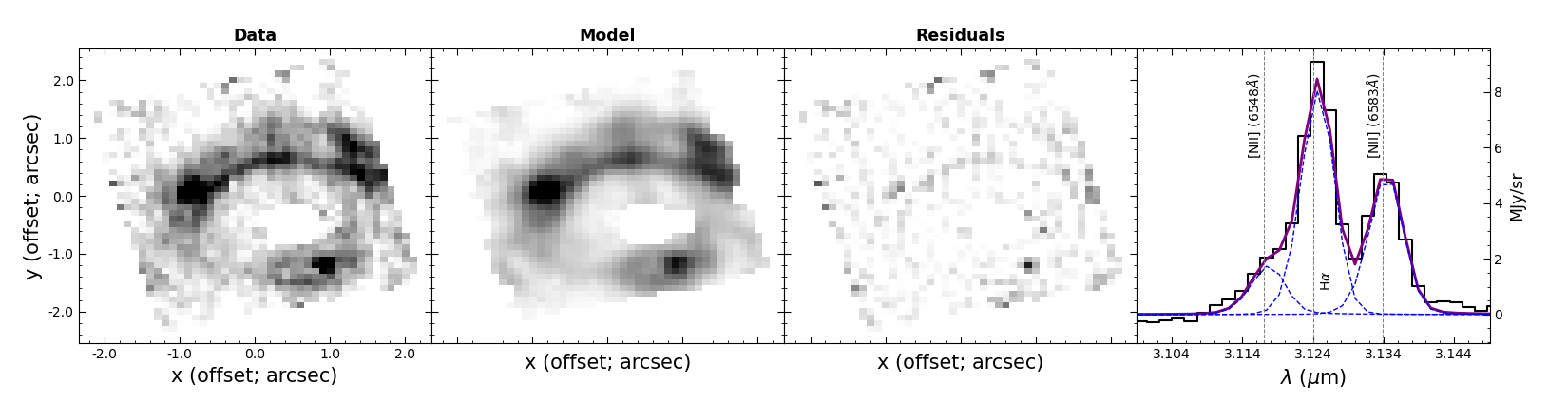}
    \caption{The result of our triple-Gaussian, pixel-by-pixel spectral fitting to the blended $\rm H\alpha$ and [N{\sc ii}] doublet, where we have assumed that the redshifts and widths of all lines are the same for all pixels. The first three panels, from left to right, show the data, the model and residuals of the velocity-integrated lines. In the final panel we show the spatially-integrated spectrum as the black histogram. The blue lines correspond to the model spectrum of each of the three individual lines, and the purple line shows their sum.}
    \label{fig:nirspec_spectral_fitting}
\end{figure*}

We make use of JWST data from different instruments -- NIRCam, MIRI and NIRSpec -- that were taken as part of the Early Release Science Program TEMPLATES (Targeting Extremely Magnified Panchromatics Lensed Arcs and Their Extended Star formation; ID 1355; PI: Jane Rigby). We use the archival data, which were reduced with the default pipeline. Some additional post-processing steps were performed for the NIRCam and MIRI data, which we describe below.

Images with NIRCam were taken in four filters, F200W, F277W, F356W and F444W, corresponding to rest-frame wavelengths between $\lambda_{\rm rest} \sim 0.42 - 0.92 \, {\rm \mu m}$. The pixel scale of the NIRCam images is 0.06 arcsec for filters F277W, F356W and F444W and 0.03 arcsec for the F200W filter. MIRI images were taken in seven filters, F560W, F770W, F1000W, F1280W, F1500W, F1800W and F2100W, corresponding to rest-frame wavelengths $\lambda_{\rm rest} \sim 1.2 - 2.7 \, {\rm \mu m}$. The pixel scale of all MIRI images is $0.1\,\mathrm{arcsec}$. In terms of pre-processing, prior to modelling our images (see Section~\ref{sec:section_3}) we removed a constant background emission independently from each filter in the NIRCam and MIRI observations. This constant background becomes apparent when plotting the distributions of all pixel fluxes for these images, which do not peak around 0. To estimate this background we fit a Gaussian function to the distribution of all pixel fluxes and subtract the best-fitting mean value from each image. We note that when fitting a Gaussian to the pixel flux distribution we considered the part of the distribution below the peak. This is to avoid including emission from real sources in the estimation of the background, which only contribute to the positive tails of these distributions.

The NIRSpec Integral Field Unit (IFU) observations were carried out using the F290LP filter, which covers a wavelength range between 2.87 and 5.27 $\mu${\rm m}, or rest-frame between 0.6 and 1.1 $\mu${\rm m}. Several emission lines were detected across the wavelength range covered by this instrument \citep[][]{2023arXiv230710412B}. Among these lines, $\rm H\alpha$ ($\lambda_{\rm rest} = 6562.819$~{\AA}) is the brightest, although it is blended with the [N{\sc ii}] doublet ($\lambda_{\rm rest} = 6548.05$~{\AA} and $6583.46$~{\AA}). For the purpose of this work we are interested in the distribution of the H$\alpha$ emission line, and so, several steps were taken to disentangle its emission.

{First, we subtract the continuum emission from the observed cube by fitting a second order polynomial to the spectrum of each pixel. When fitting the continuum, we mask the regions of the spectrum containing emission lines. Furthermore, the continuum is only being fitted in a frequency range around these emission lines. For example, given the redshift of SPT\nobreakdash-2147, the observed-frame frequency of the H$\alpha$ line is, $\lambda_{\rm obs}=3.12$ $\mu$m, and so we only fit the continuum in the frequency range, 3.0 $\mu$m < $\lambda$ < 3.3 $\mu$m (after masking the line). This latter condition is necessary since the continuum emission across the whole frequency range of the observations displays variability that a single second-order polynomial cannot capture. The following steps use the continuum-subtracted cube.}

To de-blend the emission from each individual line (i.e. H$\alpha$ + [N{\sc ii}] doublet) we fit a model consisting of three Gaussian profiles to each individual pixel of the observed cube. During optimization we allowed the centre of the Gaussian profile corresponding to the H$\alpha$ line to be a free parameter while the centres of the [N{\sc ii}] doublet were fixed according to their frequency difference with respect to H$\alpha$. In addition, we assume that the widths of all three lines are the same and so the only free parameters are the central wavelength (i.e. mean) and width (i.e. standard deviation) of the Gaussian profile fitting the H$\alpha$ emission line, and the amplitudes of each of the three individual Gaussian profiles.

In Figure~\ref{fig:nirspec_spectral_fitting} we show the results from our spectral fitting analysis, specifically the velocity-integrated intensity (i.e. 0$^{\rm th}$ moment map), where the velocity range used for the integration covers all three lines. From left to right the first three panels show the data, best-fitting model and residuals. We can already see differences between the image-plane H$\alpha$ emission distribution and that of the other tracers, suggesting different source-plane morphologies (see Section~\ref{sec:morphologies}). In the final panel of this figure we show the spatially integrated (using an elliptical annular mask) 1D spectrum for the data as the black histogram and the model as the purple line. The blue dashed lines show the one-dimensional spectra of each individual line. 

\subsubsection{HST}\label{sec:HST}

SPT\nobreakdash-2147 was observed using the Wide Field Camera (WCF3) onboard HST in the F140W filter, which samples the rest-frame ultraviolet (UV) emission ($\lambda_{\rm rest}\sim\,$0.29\,$\rm \mu m$). At these shorter wavelengths, the emission from the lens's light overwhelms the emission from the background source, making the latter imperceptible. The image is shown in Appendix~\ref{sec:Appendix_A}. 

Employing a lens light subtraction technique (see Section~\ref{sec:section_3}) we subtract the lens light from the image and identify low level emission ($\rm SNR_{\rm peak} \sim 3$) coincident with the bright arc in the NIRCam images, which we associate with the background source. Most of the emission seen with HST coincides with the part of the arc that has low dust continuum emission, which is to be expected given that dust attenuation at these wavelengths is typically severe. Unfortunately, however, the SNR of the background source emission is not high enough for us to attempt to reconstruct it in the source plane. Therefore, we will not use these data for the rest of this work. However, it is still informative to look at the HST data in comparison with the JWST data and highlight the need for longer wavelength observations of this dusty source. 

In addition, our lensed light subtraction reveals a bright compact clump (west of the bottom lensed image in Figure~\ref{fig:hst_lens_light_subtraction_subplots}), which is also seen in the bluer NIRCam filters (most prominently in the F200W filter, but also in F277W and F356W). This compact clump does not appear to have a counter image associated with it and so we conclude that it is not associated with the background source, a conclusion that is supported by their relative colours. Instead we classify this clump as a line-of-sight source most likely located in front of the lensing galaxy (see Appendix~\ref{sec:Appendix_A}).

\subsubsection{ALMA}\label{sec:data_ALMA}

SPT\nobreakdash-2147 has been observed with ALMA in multiple bands, targeting the continuum emission from the dust as well as a variety of far-infrared and sub-mm emission lines. Continuum observations were taken using different antenna configurations, leading to different resolutions. In order to best complement our data from JWST, we make use of the highest resolution ALMA continuum dataset available in the archive. These data come from proposal 2019.1.00471, and achieve a resolution of $\sim0.1\,\mathrm{arcsec}$ in band 8. 

Among the far-infrared emission lines that have been observed for this source is the [C{\sc ii}] emission line, which is the brightest far-infrared emission line and traces emission from gas in various phases (atomic/molecular/ionized). The [C{\sc ii}] emission line observations were carried out as part of proposal 2018.1.01060.S and achieve a resolution of $\sim0.25\,\mathrm{arcsec}$. As part of the same proposal the CO(7-6) and [C{\sc i}](2-1) were also observed for this source which we also use in this work, although mostly for visualization purposes.

We calibrate the data using the ALMA pipeline in the Common Astronomy Software Applications package \citep[CASA;][]{2007ASPC..376..127M}. For visualization purposes we produce cleaned images of both the dust continuum and [C{\sc ii}] line emission, which we show in the bottom left and right panels of Figure~\ref{fig:fig1}, respectively. 

\begin{table*}
    \centering
    \caption{Physical properties: stellar mass ($\rm M_{\star}$), infrared luminosity ($\rm L_{\rm IR}$), star formation rate (SFR) and age of the stellar population, derived from our SED fitting analysis with {\sc magphys}. We also report the gas mass, $\rm M_{\rm gas}$, derived from the observed CO(2-1) line luminosity, which is taken from \protect\cite{2016MNRAS.457.4406A}. All physical properties are given after correcting for magnification, which is estimated to be, $\mu = 6.4 \pm 0.5$.}
    \begin{tabular}{cccccc}
        \hline
        & $\rm L_{\rm IR}$ & ${\rm SFR}$ & $\rm M_{\star}$ & $\rm M_{\rm gas, {\rm CO(2-1)}}$ & age \\
        & ($\rm L_{\odot}$) &  ($\rm M_{\odot} yr^{-1}$) & ($\rm M_{\odot}$) & ($\rm M_{\odot}$) & (Gyr) \\
        \hline
        \textbf{SPT\nobreakdash-2147} & (9.4 $\pm$ 1.4) $\times \, 10^{12}$  & 781 $\pm$ 99 & (6.3 $\pm$ 0.9) $\times \, 10^{10}$ & (2.8 $\pm$ 0.7) $\times \, 10^{10}$ & 0.8 $\pm$ 0.1 \\
        \hline
    \end{tabular}
    \label{tab:properties}
\end{table*}

\begin{table*}
    \centering
    \caption{The best-fitting mass model parameters for SPT\nobreakdash-2147, from fitting the F444W NIRCam image using the open-source software {\sc PyAutoLens}. The definition of each parameter can be found in the main text (see Section~\ref{sec:mass_model}). Note that we report values of the axis ratio and amplitude (with their corresponding positions angles, measured counterclockwise from the positive x-axis) of the EPL and external shear models, respectively, computed from components ($e_1, e_2$) and ($\gamma_{\rm ext, 1}, \gamma_{\rm ext, 1}$).} 
    \begin{tabular}{ccccccc}
        \hline
        & $\alpha$ & $\theta_{\rm E}$ & q & $\theta$ & $\gamma$ & $\theta_{\gamma}$ \\
        & & (arcsec) & & (deg) & & (deg) \\
        \hline
        \textbf{F444W} & 2.05 $\pm$ 0.04  & 1.175 $\pm$ 0.003 & 0.85 $\pm$ 0.03 & 85 $\pm$ 4 & 0.04 $\pm$ 0.01 & 165 $\pm$ 7 \\
        \hline
    \end{tabular}
    \label{tab:mass_model_parameters}
\end{table*}

\subsection{Galaxy integrated Properties}\label{sec:properties}

In this section, we present the methods employed for estimating physical properties of the source: stellar mass, star formation rate, and molecular gas mass. The first two properties enable us to determine the location of the galaxy relative to the main sequence of star-forming galaxies \cite[e.g.,][]{2014ApJ...795..104W} and establish whether if it follows the mean trend or is an outlier (e.g. a starburst galaxy). The last quantity allows us to estimate for how long the galaxy can sustain its rate of star formation.

\begin{figure}
    \centering
    \includegraphics[width=0.995\columnwidth,height=0.25\textheight]{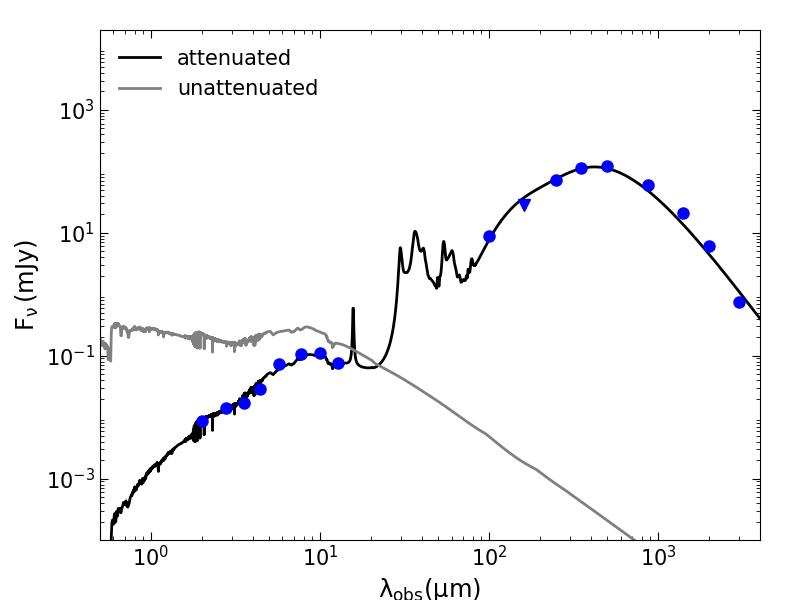}
    \caption{Spectral energy distribution fitting for SPT\nobreakdash-2147, from {\sc magphys}. The blue points correspond to the observed fluxes, taken with a variety of instruments, JWST, Herschel, LABOCA, ALMA, SPT, while the black/gray curves shows the best-fit attenuated/unattenuated model.
    }
    \label{fig:sed}
\end{figure}

\subsubsection{SED fitting}\label{sec:SED}

The typical approach for estimating the physical properties of a galaxy is to model its spectral energy distribution (SED). This is a powerful technique especially when measurements of the flux across the whole spectrum (at least from UV to far-infrared) are available. Various ingredients can be included in SED models, e.g. prescriptions for the stellar populations and dust attenuation, and different techniques have been developed \citep[e.g.][]{2015ApJ...806..110D, 2021ApJS..254...22J}.

In this work, we perform SED modelling using the high-redshift version of the publicly available software {\sc magphys} \citep[][]{2015ApJ...806..110D, 2019ApJ...882...61B}, which assumes a Chabrier IMF and employs an energy balance technique whereby all of the UV/optical emission from newly born stars is deemed to be absorbed by dust and re-emitted at infrared wavelengths. The SED fitting was carried out using fluxes from rest-frame UV/optical ($\lambda_{\rm rest, min} \sim 0.42$ $\mu$m) to sub-mm/mm wavelengths ($\lambda_{\rm rest, max} \sim 630$ $\mu$m). If only upper limits exist in some band these were not used in the fitting process (the only upper limit we have is for the flux density at $160\,\mathrm{\mu m}$ and is consistent with the best-fitting solution plotted in Figure~\ref{fig:sed}).

We note that the SED fitting was performed prior to correcting the observed fluxes for magnification, since some of our observations, specifically in the sub-mm regime (from Herschel, LABOCA and SPT), do not resolve the source and therefore cannot be reconstructed (i.e. corrected for magnification). However, the average magnification factor that we measure using stellar continuum observations from JWST, specifically in the F444W filter image, is in agreement with the value measured using dust continuum observations from ALMA, so we expect that our results are not significantly affected by differential magnification (i.e. different parts of the source experiencing different magnification factors).

In Figure~\ref{fig:sed} we show the results from our SED fitting analysis. The blue points correspond to our observed fluxes while the black line corresponds to the best-fitting model. The best-fit model provides a good fit to the data across the whole spectrum, with a overall goodness of fit, $\chi^2 \sim 3$. The physical properties of our source derived from this best-fitting model are summarized in Table~\ref{tab:properties}: stellar mass ($\rm M_{\star}$), far-infrared luminosity ($\rm L_{\rm IR}$), star formation rate (SFR), and age of the stellar population. The values that we estimate for the $\rm L_{\rm IR}$, SFR and $\rm M_{\star}$ are in good agreement with previous measurements for this source \citep[][]{2016MNRAS.457.4406A, 2022A&A...663A..22G, 2023ApJ...958...64B}. {
From the measured values of the SFR and $\rm M_{\star}$ we conclude that SPT-2147 is a starburst galaxy. It's SFR is more that $\times 4$ above the main-sequence of star-formation galaxies at z=3.78 \citep[][]{2023MNRAS.519.1526P}.}

\subsubsection{Gas mass}

Molecular gas is the raw fuel for the formation of new stars in a galaxy. To measure this quantity several approaches have been proposed. Ideally one would like to measure the amount of molecular hydrogen ($\rm H_2$) in a galaxy directly, but this is observationally challenging for objects at $z >$ 0.1. Alternatively, one can rely on different tracers for estimating molecular gas masses, with the most frequently used being carbon monoxide (CO), the second most abundant molecule in a galaxy after $\rm H_2$. Typically, the ground-state transition, CO(1$-$0), is used in this approach, which assumes a conversion factor \citep[$\alpha_{\rm CO}$;][]{2013ARA&A..51..207B} to compute the molecular gas mass from the CO line luminosity, $\rm L_{\rm CO}$.


Unfortunately, in our case we do not have a measurement for the CO(1$-$0) emission line flux, but, as mentioned in Section~\ref{sec:data_ALMA}, observations of the CO(2$-$1) emission line have previously been obtained for SPT\nobreakdash-2147 \citep[][]{2016MNRAS.457.4406A, 2022A&A...663A..22G}. Here we use the CO(2$-$1) emission line luminosity, $L_{\rm CO(2-1)} = (1.6 \pm 0.3) \times 10^{11} {\rm (K \, km \, s^{-1} \, pc^{2})}$, to estimate the molecular gas mass from
\begin{equation}\label{eq:eq_gas_mass}
    M_{\rm gas} = 1.36 \, \alpha_{\rm CO} \, r_{21} \, L_{\rm CO(2-1)},
\end{equation}
where the parameter $r_{21}$ represents the line ratio of the CO(2-1) transition to the ground state transition, for which we adopt the value $r_{21} = 0.84 \pm 0.13$ \citep[e.g.][]{2013MNRAS.429.3047B} and $\alpha_{\rm CO}$ is the so-called $\rm CO-H_2$ conversion factor, given in units of $\mathrm{M}_{\odot}\,\mathrm{K}^{-1}\,\mathrm{km}^{-1}\,\mathrm{s}\, \mathrm{pc}^{-2}$, for which we adopt the value, $\alpha_{\rm CO} = 1.0$ \citep[e.g.][]{2021MNRAS.501.3926B}. The factor of 1.36 accounts for the abundance of Helium. We estimate a molecular gas mass, prior to correcting for magnification, of $M_{\rm gas} = (1.8 \pm 0.4) \times 10^{11}$~${\rm M_{\odot}}$. With both the stellar and gas masses in hand we can measure the gas fraction for SPT\nobreakdash-2147 (ratio of gas to stellar mass) for which we find, $\rm f_{\rm gas} = 0.45 \pm 0.12$, which is on the high end but consistent with previous estimates for SMGs/DSFGs \citep[][]{2021MNRAS.501.3926B}.

\section{Strong Lensing Analysis} \label{sec:section_3}

In this section, we give a general description of the methodology and models that we used to obtain source-plane reconstructions. We perform lens modelling (which includes lens light subtraction, lens mass model optimization and background source reconstructions) using the open-source software {\sc PyAutoLens}\footnote{\url{https://github.com/Jammy2211/PyAutoLens}}, which is described in \citet{2015MNRAS.452.2940N, 2018MNRAS.478.4738N, 2021JOSS....6.2825N} and builds on methods from previous works in the literature \citep[e.g.][]{2003ApJ...590..673W, 2006MNRAS.371..983S, 2009MNRAS.392..945V}. 

The datasets that we model are either the conventional CCD images (e.g. NIRCam, MIRI) in which case the model {\sc PyAutoLens} produces is also an image that attempts to match the data on a pixel-by-pixel basis, or visibilities \citep{2017isra.book.....T} which are the basic product that interferometers, such as ALMA, record. In the latter case one can produce `dirty images' from these visibilities and carry out the analysis as one would do for imaging data \citep[e.g.][]{2018MNRAS.476.4383D} or choose to work directly in the so-called $uv$-space by modelling the visibilities \citep[e.g.][]{2018MNRAS.475.3467E, 2021MNRAS.501..515P, 2022MNRAS.512.2426M}. When reconstructing the dust continuum and [C{\sc ii}] line emission distributions with ALMA we make use of the latter approach.

In order to build a mass model for the lensing galaxy we choose the F444W filter image from JWST/NIRCam. The reasons for choosing to build the model from the NIRCam image rather than the ALMA dataset is that the resolution is about twice as high and the structure of the background source appears more complex, which generally leads to better constraints for the parameters of the mass model \footnote{We independently modelled the dust continuum emission from high-resolution ALMA observations and found that the best-fitting mass model parameters are in good agreement with those derived from the F444W NIRCam image. A detailed description on the modelling of the intereferometric data from ALMA is beyond the scope of this paper.}. As for choosing the F444W filter out of four filter images observed with NIRCam, the reason is twofold. Firstly, at these longer wavelengths the emission from the lens is weaker compared to the other filters, which allows for a cleaner subtraction of its light distribution and therefore minimizes potential contamination of the lensed emission \citep[][]{2022arXiv220910566N}. Secondly, the SNR of the lensed emission from the background source is higher at this wavelength, whereas bluer wavelengths are generally more affected by dust obscuration, resulting in better constraints for the mass model parameters. 

The analysis is carried out in several phases, whereby succesive phases introduce more complexity to either the mass and/or the source model and inherit information, in the form of priors, from previous phases. The different phases of the pipeline that {\sc PyAutoLens} goes through to build a mass model for the lensing galaxy have been described in previous papers that make use of the same software \citep[e.g.][]{2022MNRAS.517.3275E, 2022RAA....22b5014C}, and are the same steps that we take in this work. In the sections that follow we describe the profiles that we used to model the lens light and mass distributions. 

\begin{figure*}
    \centering
    \includegraphics[width=0.95\textwidth, height=0.9\textheight]{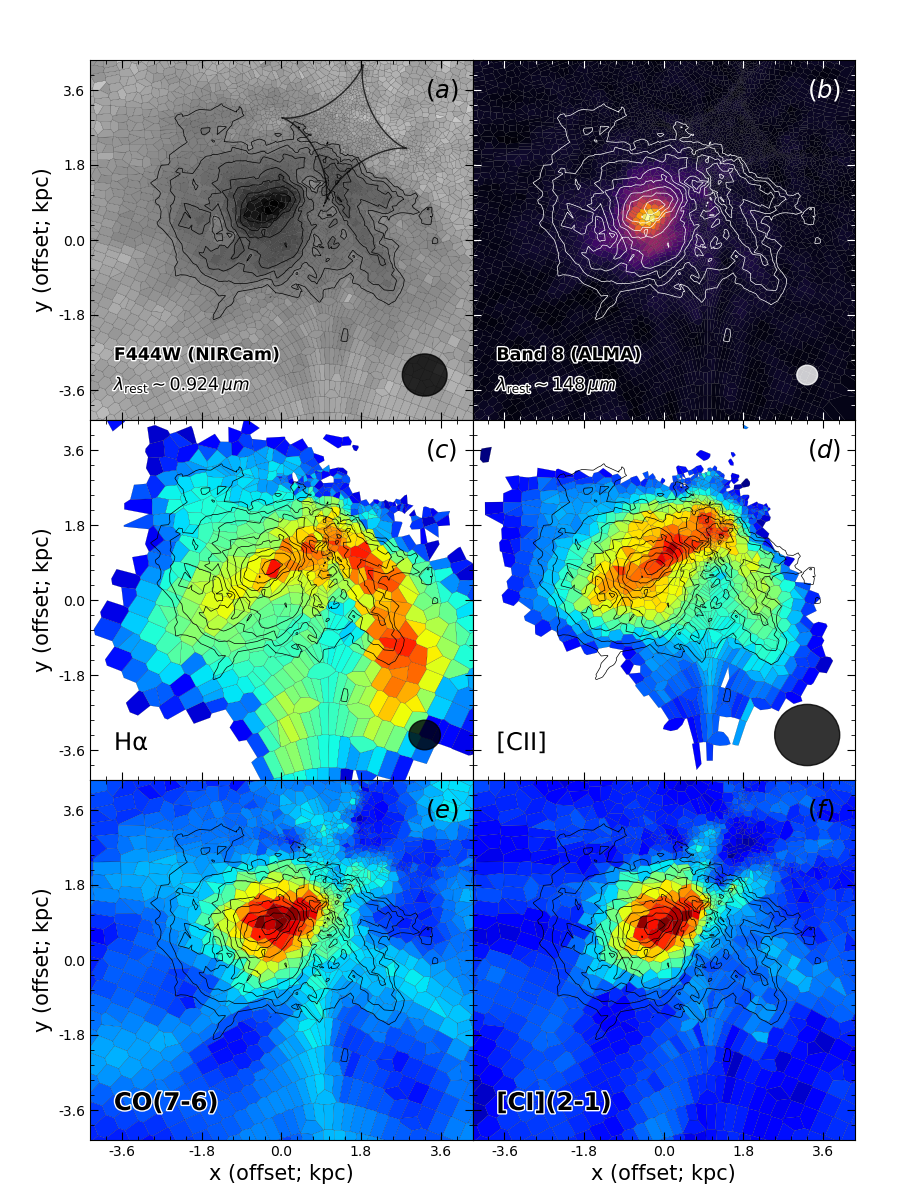} 
    \caption{Source-plane reconstructions for SPT\nobreakdash-2147 of the stellar continuum (F444W NIRCam filter corresponding to rest wavelength of $\lambda_{\rm rest} \sim 0.92 \, {\rm \mu m}$; top left), dust continuum (band 8, $\lambda_{\rm rest} \sim 148 \mu {\rm m}$; top right), the velocity-integrated intensity ($0^{\rm th}$ moment) of the $\rm H\alpha$ and [C{\sc ii}] emission lines (middle left and right, respectively) and of the CO(7-6) and [C{\sc i}](2-1) lines (bottom left and right, respectively). Contours of the stellar continuum emission are superimposed in all panels to help visualize the spatial relation of the different ISM components. The black curve in the top left panel shows the caustic curve. The circles in the bottom right corner of each panel denote the average resolution achieved in the source plane.}
    \label{fig:source_plane_reconstructions}
\end{figure*}

\begin{figure*}
    \centering
    \includegraphics[width=0.995\textwidth,height=0.2125\textheight]{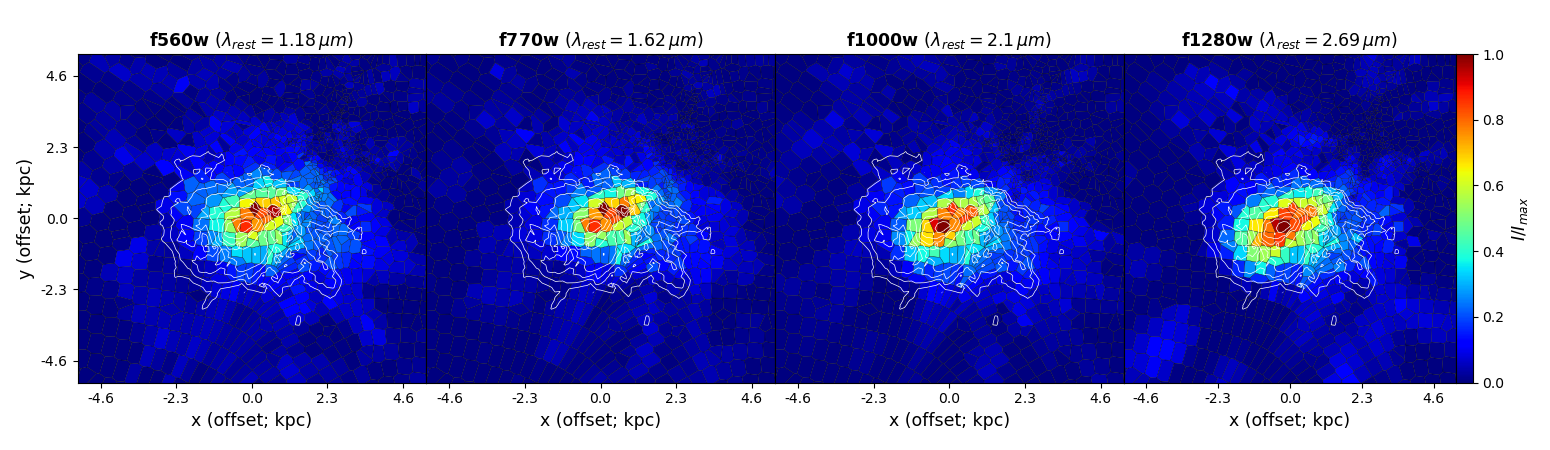} 
    \caption{Source-plane reconstructions from the MIRI images, corresponding to rest-frame wavelengths from $\lambda_{\rm rest} \sim 1.0 - 2.5 \, {\rm (\mu m)}$. The white contours in each panel correspond to the F444W reconstructions and are the same as in Figure~\ref{fig:source_plane_reconstructions}.}
    \label{fig:miri_reconstructions}
\end{figure*}

\subsection{Lens light}\label{sec:lens_light}

We used two models for the len's light emission profile. The most common type of profile used to describe the light distribution of galaxies is the S\'ersic profile \citep{1963BAAA....6...41S}. This has a functional form given by:
\begin{equation}
I(r)=I_e \exp \left\{-b_n\left[\left(\frac{r}{r_e}\right)^{1 / n}-1\right]\right\} \, ,
\end{equation}
where $r_{\rm e}$ is the effective radius, $I_{\rm e}$ the intensity at the effective radius, $n$ the S\'{e}rsic index and $b_{\rm n}$ a constant parameter that only depends on $n$ \citep{2004AJ....127.1917T}. The typical approach is actually the use of a double S\'{e}rsic profile as to model the different slopes of the light distribution observed in the centers and outskirts of galaxies \citep[e.g.][]{2019MNRAS.489.2049N}.

Despite the popularity of the S\'{e}rsic profile \citep[e.g.][]{2022MNRAS.517.3275E}, it has been noted that perhaps it lacks the necessary complexity to describe galaxy light distributions \citep[e.g.][]{2022arXiv220910566N, 2023MNRAS.518..220H}, especially when deeper observations are available. A more flexible type of profile, which can capture non-axisymmetric features is the multiple Gaussian expansion (MGE) \citep{2002MNRAS.333..400C}, which is widely used in studies of galaxy dynamics \citep[][]{2008MNRAS.390...71C}, and more recently has been applied in lensing studies as well \citep{2023MNRAS.518..220H}. 

\subsection{Foreground mass \& background source}\label{sec:mass_model}

For the mass distribution of the lensing galaxy we consider an elliptical power-law model (EPL) with an external shear. The convergence of the EPL model is defined as,
\begin{equation}\label{equ:pl_mass}
    \kappa(x, y) = \frac{3 - \alpha}{1 + q} \left( \frac{\theta_{\rm E}}{\sqrt{x^2 + y^2 / q^2}} \right)^{\alpha-1} \,
\end{equation}
where $\alpha$ is the 3D logarithmic slope, $q$ is the axis ratio (minor to major axis) and $\theta_{\rm E}$ is the Einstein radius. Our model also has additional free parameters that control the position of the centre ($x_{\rm c}, \, y_{\rm c}$) and its position angle, $\theta$, which is measured counterclockwise from the positive x-axis. In practice we parametrize the axis ratio, $q$, and position angle, $\theta$, in terms of two components of ellipticity:
\begin{equation}
    e_1 = \frac{1 - q}{1 + q} {\rm sin}\left(2\theta\right) \,; \,\,\,\,\,
    e_2 = \frac{1 - q}{1 + q} {\rm cos}\left(2\theta\right) \, .
\end{equation}
This helps to prevent periodic boundaries and discontinuities in parameter space associated with the position angle and axis ratio. Finally, our model has an additional component to model the presence of external shear. This is typically described by two parameters: a magnitude, $\gamma_{\rm ext}$, and a position angle, $\theta_{\rm ext}$, but here we express them in terms of $\gamma_{\rm ext, 1}$ and $\gamma_{\rm ext, 2}$ as:
\begin{equation}
    \gamma_{\rm ext} = \sqrt{\gamma^2_{\rm ext, 1} + \gamma^2_{\rm ext, 2}} \, , \,\,\,\,\,
    {\rm tan}2\theta_{\rm ext} = \frac{\gamma_{\rm ext, 2}}{\gamma_{\rm ext, 1}} \, .
\end{equation}

As we discussed in Section~\ref{sec:HST}, we identify a compact source southwest of the lensing galaxy. We tried including an additional component in the mass model, specifically a spherical Singular Isothermal Sphere (SIS; setting $q = 1.0$ and $\alpha = 2.0$ in Eq.~\ref{equ:pl_mass}), in order to account for its contribution to the lensing. We found, however, that when this additional component is included in the mass model, its Einstein radius, $\theta_{\rm E}$, converges to a very small value, as if its contribution is negligible. Therefore, for the remainder of this work we ignore the presence of this compact source; our results remain unchanged whether or not we include it in the mass model.

\begin{figure*}
    \centering
    \begin{tabular}{c}
    \includegraphics[width=0.95\textwidth,height=0.225\textheight]{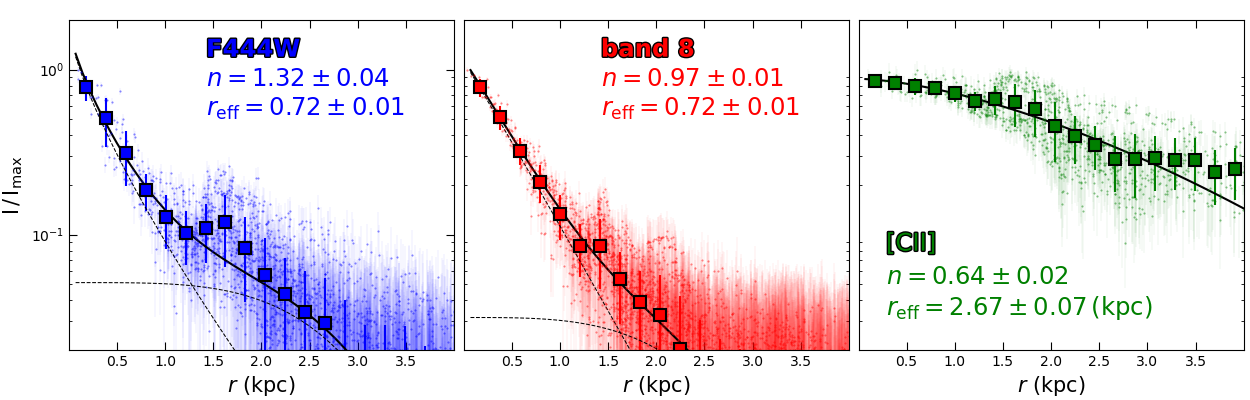} 
    \end{tabular}
    \caption{Source-plane reconstructions plotted as a function of distance from the centre for each component of the galaxy (stars, dust and gas), as indicated in the legend of each panel. The larger square symbols show the averaged profile in bins of distance; we consider both the scatter in each bin as well as the errors on individual points {(i.e. the individual resolution elements of the source reconstruvtion)} when calculating the errors. The black curves show the best-fitting single/double S\'{e}rsic model to the data. We quote the best-fitting S\'{e}rsic index, $n$, and effective radius, $r_{\rm e}$, in each panel of the figure (if a double S\'{e}rsic model was used we only show the values of S\'{e}rsic that captures the central emission). We note that SPT\nobreakdash-2147 shows an excess (or `bump') in its radial profiles at a radius, $r \sim 1.5$ kpc, which we associate with the spiral-arm like feature in our source reconstruction (see Figure~\ref{fig:source_plane_reconstructions}), and for that reason, we did not use points around that region during the fitting process. Overall, we find that the best-fitting S\'{e}rsic parameter values are consistent with our source having disc-like characteristics.}
    \label{fig:source_plane_reconstructions_1d}
\end{figure*}

\subsubsection{Background source}\label{sec:background_source}

The lensing-corrected source surface brightness distribution is reconstructed on an irregular Voronoi
mesh \citep[e.g.][]{2015MNRAS.452.2940N}. Such a mesh has the advantage that it adapts to the magnification pattern in the source plane, so that more source-plane pixels are dedicated to regions with higher magnifications (since the resolution will also be higher in these regions, scaling roughly as $\mu^{-1/2}$) compared to regions with lower magnification. When reconstructing a source on a pixelated mesh it is also necessary to introduce some form of regularization in order to avoid over-fitting \citep[e.g.][]{2006MNRAS.371..983S}. In this work we use a constant regularization scheme, where the level of regularization is controlled by the coefficient, $\lambda_{\rm reg}$. We note that this non-parametric approach, which is considered the standard in the field, is necessary to obtain accurate reconstructions as our source's morphology is complex and cannot be described with simple parametric profiles.

\subsubsection{Pipeline}
{
Our modeling pipeline is divided into several phases, with each subsequent phase introducing more complexity to either the mass distribution or the background source, inheriting from the previous phase in the form of priors.

In the initial phase of our pipeline, we fit an SIS + shear profile for the mass distribution of the lensing galaxy and a single S\'{e}rsic profile for the background source. A S\'{e}rsic profile is also used to fit the lens galaxy light when fitting imaging data; for interferometric data, light from the lens is not detected. This initial phase allows us to obtain a decent model for the lensing galaxy's mass. Subsequent phases then use a pixelated model for the background source, as described in Section~\ref{sec:background_source}.

A schematic presentation of the various phases of our pipeline is given in \cite{2022RAA....22b5014C}. The same phases are also used in this work, with the only exception being that the final phase of our imaging pipeline uses an MGE model for the lens galaxy light instead of a S\'{e}rsic profile \citep{2024arXiv240316253H}. The lens light subtracted images shown in the top right panel of Figure~\ref{fig:fig1} were produced using an MGE model for the lensing galaxy's light.
}

\subsubsection{Lens modelling systematics}
{
In the field of strong gravitational lensing, it is a well-known fact that the mass distribution of early-type galaxies is not perfectly described by an EPL model. For example, it is increasingly evident that the mass distribution is not always axially symmetric, but rather exhibits angular complexity \citep[e.g.][]{2022MNRAS.516.1808P, 2022arXiv220910566N, 2023MNRAS.518..220H}. A valid question to ask is whether the lack of this type of complexity in the mass model could affect the source reconstructions by introducing artifacts. In a recent study by \cite{2024arXiv240304850S} using high-resolution data, the authors demonstrated that these corrections are at a few percent level, and accounting for them does not alter the inferred morphology of the background sources. Therefore, we argue that even though we do not incorporate this level of complexity into our mass model, our conclusions regarding the structure of the background source should remain largely unaffected.

Another type of systematic in strong lensing modeling is the degeneracy between the slope of the power-law model and the size of the reconstructed source. For example, a steeper slope than the true slope will result in a more extended source, and conversely, a shallower slope will yield a more compact source \citep[e.g.][]{2015MNRAS.452.2940N}. This type of systematic can lead to a biased estimate of the source's size. However, since all datasets are reconstructed using the same mass model, this is not expected to introduce artifacts in the reconstructions.
}

\section{Results} \label{sec:section_4}

In this section we present reconstructions of all the available components of the galaxy (stars, dust and gas) using a fixed mass model for the lensing galaxy, whose parameters are given in Table~\ref{tab:mass_model_parameters}. In addition to reconstructing the different tracers of the ISM we also produce reconstructions of the velocity and velocity dispersion maps from the [C{\sc ii}] emission line, which we use to study the kinematics of our source. {Combining all available datasets we build a complete picture of the physical mechanisms that drive galaxy evolution in this $z \sim 3.7$ massive dusty star-forming galaxy.}

\subsection{Morphologies}\label{sec:morphologies}

We start by discussing the morphologies of the different components of the ISM. In Figure~\ref{fig:source_plane_reconstructions} we show reconstructions of the stellar continuum in filter F444W of the NIRCam instrument (top left; $\lambda_{\rm rest} \sim 0.924$ $\mu$m), dust continumm in band 8 (top right; $\lambda_{\rm rest} \sim 148$ $\mu$m), H$\alpha$ emission line (middle left; tracing ionized gas / low-obscuration star formation), [C{\sc ii}] line emission (middle right; tracing various phases of the gas -- atomic/molecular/ionized), CO(7-6) emission line (bottom left; tracing dense warm gas) and [C{\sc i}](2-1) emission line (bottom right; tracing molecular gas). 

Before we attempt to make any quantitative statements about the morphological/structural properties of our source we present the general picture that we can infer from our reconstructions. SPT\nobreakdash-2147 is an extended {($\sim 8 \,{\rm kpc}$ in diameter; the major-axis of the outermost isophote; see Section~\ref{sec:section_5})}, isolated disc galaxy (with no evidence for a companion galaxy at the current depth), viewed at an inclination of $i \sim 49 \, {\rm deg}$ (see Section~\ref{sec:dyn_mod}). The band~8 dust continuum emission appears to be compact and centrally concentrated, which is typical for this class of galaxies \citep[e.g.][]{2015ApJ...799...81S, 2016ApJ...833..103H, 2019MNRAS.490.4956G}, and suggests that most of the star formation activity is taking place at the centre of the galaxy. We find evidence of a spiral arm-like structure that is clearly detectable in the $\rm H\alpha$ emission line distribution but also becomes visible in the stellar continuum distribution, specifically the base of the spiral arm which is also seen in the stars. This spiral arm leads to an elongated boxy structure in the [C{\sc ii}] emission line distribution. The stellar continuum distribution also appears to be elongated in the centre and has the same orientation as the boxy structure in the gas. { The $\rm H\alpha$ emission appears asymmetric toward the side of the bar associated with the spiral-arm-like structure. Similar features in the morphology of the $\rm H\alpha$ distribution are also observed in local barred galaxies and can be attributed to stochastic star formation \citep[e.g.][]{2019A&A...627A..26N}.} We find that very little dust continuum emission is associated with the arm-like features, even though a previous study has suggested such features are visible in this component of the ISM \citep[][]{2019ApJ...876..130H} on similar scales. Also, very little H$\alpha$ emission is associated with the centre of the galaxy, perhaps due to the high dust content which can heavily attenuate the emission from this line. 

We also produce reconstructions from the MIRI images, which we show in Figure~\ref{fig:miri_reconstructions}. The MIRI images trace deeper into the rest-frame infrared part of the spectrum, $\lambda_{\rm rest} = 1.0 - 2.5 \, {\rm \mu m}$, and therefore suffer less from dust attenuation. However, the resolution is coarser compared to the NIRCam reconstruction and so we do not see any detailed structures in these reconstructions. What we do see is that the morphologies of the MIRI reconstructions appear to be elongated along the same direction as the linear structure that we identify in the gas and F444W images.

Considering all of the evidence above, we suggest that this source is a disc galaxy with a bar and spiral arms. If our interpretation is correct then this would be the second highest redshift galaxy where such features have been observed to date \citep[][]{2023arXiv230616039S}. We discuss the implications of this discovery in Section~\ref{sec:section_5}.

\subsubsection{1D profiles}\label{sec:1d_morphologies}

We first analyze the one-dimensional profiles of our reconstructions (flux versus distance from the centre); these are displayed in Figure~\ref{fig:source_plane_reconstructions_1d}, which shows from left to right, the one-dimensional profiles of the stellar continuum (F444W/NIRCam), band 8 dust continuum and [C{\sc ii}] emission line. The small circular points in each panel of this figure show the flux of each individual reconstructed pixel as a function of distance from the common centre\footnote{Both the stellar and dust continuum distribution peak at the same location, which we used as the common centre for all reconstructions. We note, however, that the peak in the [C{\sc ii}] emission line distribution is slightly offset compared to these.}, and square are the averaged profiles, in radial bins. One notable feature in all of the 1D profiles is a bump around $r \sim$1.5 kpc. Such a feature is often seen in the averaged profiles of spiral galaxies in the local Universe and so we attribute this feature to the presence of spiral-arms in our galaxy.

We fit a single/double S\'{e}rsic model to the average 1D profile; the best-fitting models are shown as the black solid curve in each panel of Figure~\ref{fig:source_plane_reconstructions_1d} (if a double S\'{e}rsic model is used then the individual S\'{e}rsic models are shown as the black dotted lines). The stellar and dust profiles prefer a double S\'{e}rsic model, as a single S\'{e}rsic model is not able to capture the emission close to the centre of the galaxy as well as the more extended emission in the outskirts (i.e. the extended disc). We did not attempt to fit a model to the 1D profile of the H$\alpha$ emission, as it appears to be flattened at radii of $r \geq 2$ kpc, which coincides with the outer radius of the bump feature in the 1D profile of the stellar distribution, beyond which the error bars become rather large.

For the stellar distribution, the best-fitting values of the S\'{e}rsic index and effective radius are $n = 1.32 \pm 0.04$ and $r_{\rm e} = 0.72 \pm 0.01$~kpc for the central component of the double S\'{e}rsic model, and $n = 0.25 \pm 0.01$ and $r_{\rm e} = 1.96 \pm 0.03$~kpc for the extended component. The equivalent values for the dust continuum distribution are $n = 0.97 \pm 0.01$ and $r_{\rm e} = 0.72 \pm 0.01$~kpc for the central, and $n = 0.29 \pm 0.04$ and $r_{\rm e} = 1.80 \pm 0.04$~kpc for the extended components. A S\'{e}rsic index of $n = 1$ is typically associated with disc galaxies and is what many previous studies have found when studying the resolved morphologies of the dust continuum emission \citep[e.g.][]{2016ApJ...833..103H, 2019MNRAS.490.4956G}. The slope of the central component for the stellar distribution is slightly steeper but still within the range found for disc galaxies (a S\'{e}rsic index $n = 2$ could be considered evidence for the presence of a pseudo-bulge). Finally, the 1D profile of the [C{\sc ii}] emission, which prefers a single S\'{e}rsic component, has $n = 0.64 \pm 0.02$ and $r_{\rm e} = 2.67 \pm 0.07$~kpc. 

\begin{figure}
    \centering
    \includegraphics[width=0.95\columnwidth,height=0.25\textheight]{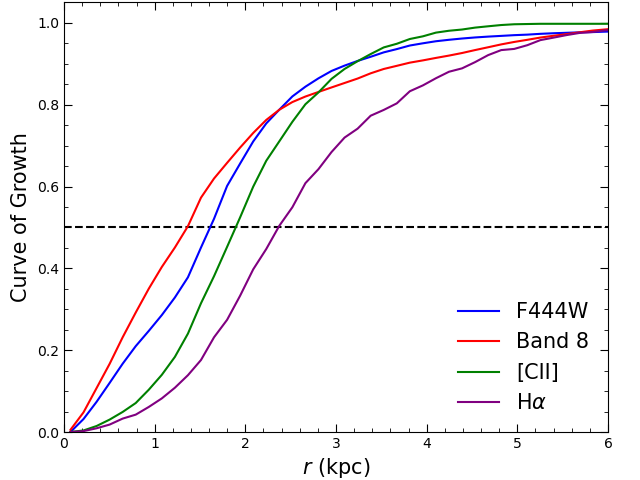} 
    \caption{Cumulative fraction of the integrated flux (i.e. curve of growth) for all the different tracers of the ISM in SPT\nobreakdash-2147, where colours are the same as in Figure~\ref{fig:source_plane_reconstructions_1d} with the addition of the H$\alpha$ emission line in purple. The radius where the lines intercept 50~per~cent of the flux corresponds to the half-light radius and can be used as an estimate of size of those components: $\sim 1.4$~kpc (Band 8), $\sim 1.6$~kpc (F444W), $\sim 1.9$~kpc ([C{\sc ii}]), $\sim 2.4$~kpc (H$\alpha$).}
    \label{fig:COG}
\end{figure}

A useful comparison for the spatial extent of the emission can be made from measuring the cumulative fraction of the flux (i.e. the curve of growth) as a function of distance from a common centre, for the different components of the galaxy. This approach does not rely on any assumptions for how the distributions are parameterized (i.e. S\'{e}rsic profiles), which is more informative for measuring sizes when the distributions are complicated, as is the case here. 

We show the curves of growth for each of the different ISM tracers in Figure~\ref{fig:COG}. We can use this to estimate the half-light radius of each component. The sizes measured from the curves of growth are consistent with our findings from the 1D S\'{e}rsic profiles. In this case we can also compute a size for the H$\alpha$ emission line distribution which appears to be the most extended of all of the components. Another piece of information that we can extract from this measurement is that $\sim$20~per~cent of the dust continuum emission is associated with an extended component, perhaps implying that some of the dust is co-spatial with the arm feature that we see in our reconstructions. 

{
Using the above structural properties for SPT\nobreakdash-2147 together with its physical properties discussed in Section~\ref{sec:properties}, we can compare our galaxy to the population of star-forming galaxies at high redshift in terms of its size. Adopting the effective radius of the extended component, from fitting to the F444W reconstructed distribution, as the galaxy's size, we infer that SPT\nobreakdash-2147 sits above the size-mass relation for star-forming galaxies \citep[][]{2024ApJ...962..176W}.
}

\subsection{Kinematics}

\begin{figure*}
    \centering
    \includegraphics[width=0.995\textwidth,height=0.35\textheight]{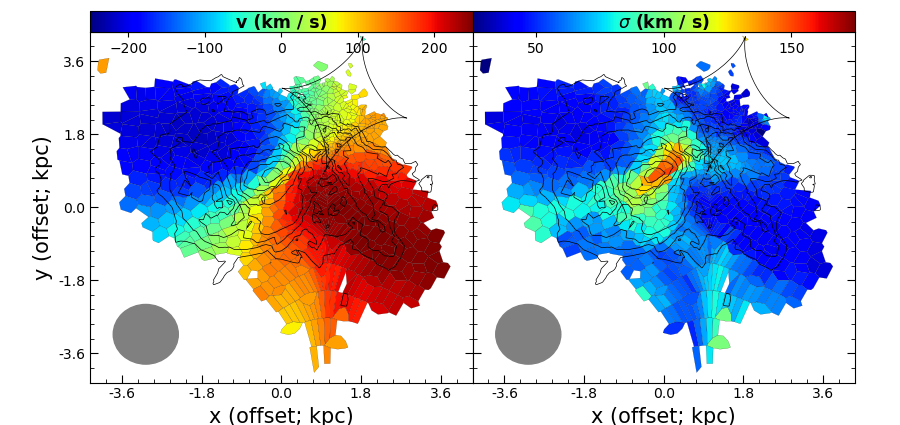} \\

    \caption{Source-plane reconstructions of the 1$^{\rm st}$ (velocity; left panel) and 2$^{\rm nd}$ (velocity dispersion; right panel) moment maps of the  [C{\sc ii}] emission line. The black curves in each panel show the caustic curve while the black contours are of the stellar continuum emission in the F444W NIRCam filter (same as Figure~\ref{fig:source_plane_reconstructions}). The gray circle at the bottom left corner of each panel represents the beam in the source plane, which assumes that the resolution scales as $\mu^{-1/2}$, resulting in $\sim$20 independent resolution elements across the area of the source.}
    \label{fig:source_plane_mom_1_reconstructions}
\end{figure*}

Next, we turn our attention to the gas kinematics of our source by modelling the reconstructed [C{\sc ii}] emission line cube. The reason for choosing the [C{\sc ii}] over the H$\alpha$ emission line is two fold: it has a higher signal-to-noise ratio, providing better constraints to the dynamical model parameters, and was observed at higher velocity resolution, $\sim 6$ $\rm km \, s^{-1}$ instead of $\sim 100$ $\rm km \, s^{-1}$ for the H$\alpha$ cube, which is necessary for constraining the velocity dispersion. Before we dive into the details of the dynamical modelling analysis, in Figure~\ref{fig:source_plane_mom_1_reconstructions} we show the source-plane reconstructions of the 1$^{\rm st}$ and 2$^{\rm nd}$ moment maps. 

The reconstructed velocity and velocity-dispersion maps of the [C{\sc ii}] emission line distribution were created by first reconstructing (in the $uv$-plane) each individual channel of the emission line cube, where we binned the emission in channels of velocity width $\sim 30$ $\rm km \, s^{-1}$ to further increase the signal-to-noise ratio. With the reconstructed cube we fit the spectrum of each reconstructed Voronoi cell with a Gaussian profile (similar to the approach we followed in Section~\ref{sec:section_2}). The best-fitting values of the mean and standard deviation of the Gaussian profile in each pixel correspond to the velocity and velocity dispersion values, respectively, for that cell. 

In order to determine whether the Gaussian profile is a better fit to the spectrum in each cell, we compared its goodness-of-fit (i.e. $\chi^2$) to that of a straight line going through the abscissa (i.e. signal consistent with noise). We define the parameter, ${\rm SNR}^2_{\rm G} = \chi^2_{\rm G} - \chi^2_{\rm line}$, which is the difference of $\chi^2$ values for a Gaussian profile and that of a straight line. For our reconstructed moment maps in Figure~\ref{fig:source_plane_mom_1_reconstructions} we are only showing pixels for which ${\rm SNR}_{\rm G} > 5$.

As Figure~\ref{fig:source_plane_mom_1_reconstructions} shows, the reconstructed moment maps that our source displays characteristics of a rotating disc galaxy. There is a well-defined velocity gradient along the major axis of the source covered by enough resolution elements (the beam is shown as a gray circle in the bottom left corner of each panel) so that it is not a consequence of beam smearing (we have $\sim$ 20 independent resolution elements across the area of the disc). The velocity dispersion map shows a distinct peak which coincides with the centre of the stellar distribution (the black contours on each are the same as in Figure~\ref{fig:source_plane_reconstructions}). In the next section, we perform a 3D kinematic modelling analysis in order to constrain parameters that quantify the dynamical state of our source.

\subsubsection{Dynamical modelling}\label{sec:dyn_mod}

\begin{figure*}
    \centering
    \begin{tabular}{c}
    \includegraphics[width=0.995\textwidth,height=0.275\textheight]{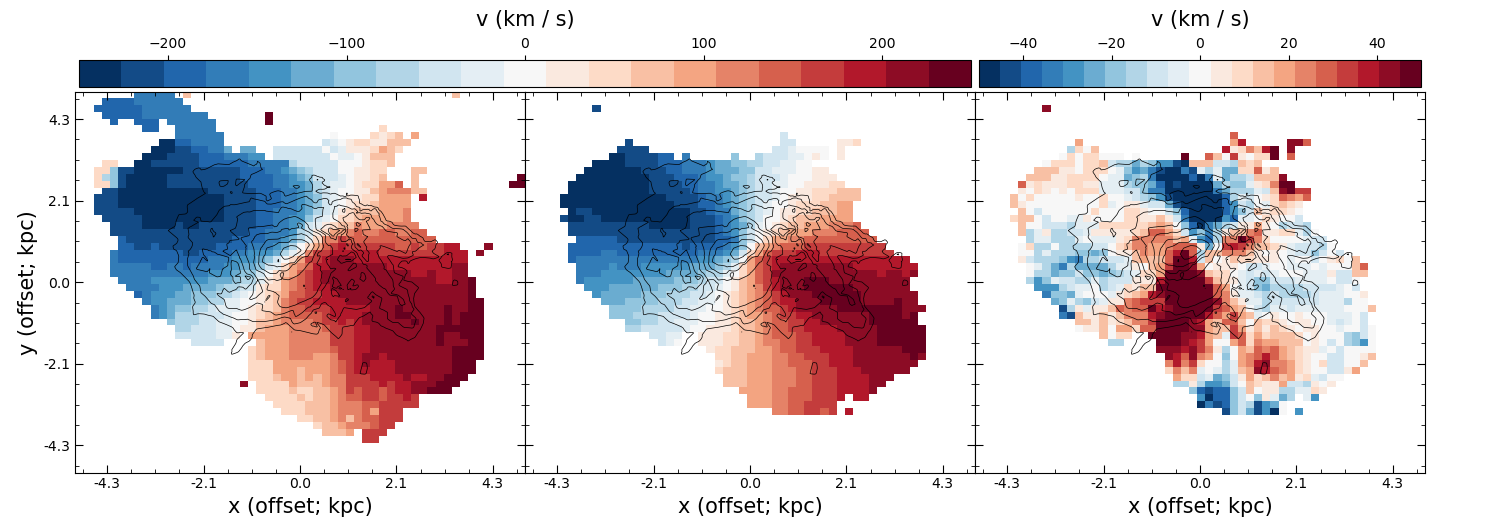} \\
    \includegraphics[width=0.995\textwidth,height=0.4\textheight]{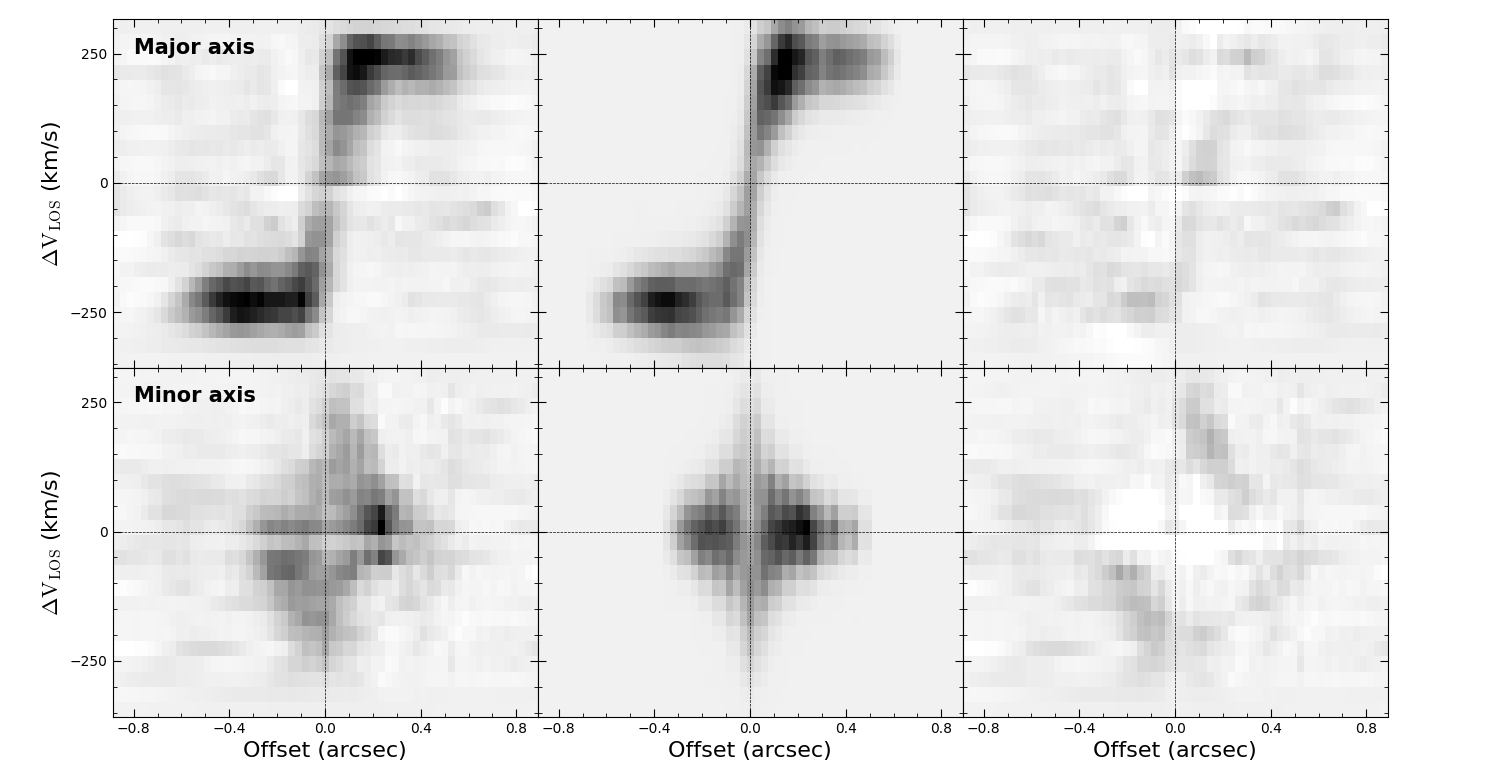} \\
    \end{tabular}
    \caption{In the top row we show the data, model and residuals for the reconstructed velocity map. There are some residuals left over by the smooth rotating disc model, which coincide with the leading edges of the bar, that could be explained by the presence of non-circular motions along the bar in this galaxy (i.e. in-flowing gas to the centre). Note that the differences in the velocity map in here and Figure~\ref{fig:source_plane_mom_1_reconstructions} are due to the reconstruction having been interpolated onto a regular grid. In the middle and bottom rows we show the position-velocity diagrams across the major and minor axis, respectively, for SPT\nobreakdash-2147. The different panels from left to right show the data, model and residuals, all shown on the same colorscale.
    }
    \label{fig:dynamical_modelling_figure_1}
\end{figure*}

In order to model the dynamics of SPT\nobreakdash-2147 we use the publicly available software { $^{\rm \sc 3D}${\sc Barolo}}\footnote{Version 1.6.9; git commit ID 171b30b379fdc76e402cb0c8cf6d8cbb8e784196.} \citep[][]{2015MNRAS.451.3021D}. The main function of this code is to iteratively fit a 3D tilted-ring model to an observed data cube (in our case reconstructed from the observed lensed emission), optimizing for each ring the inclination, position angle, systemic velocity ($\rm V_{\rm sys}$), rotational velocity ($\rm V_{\rm rot}$) and velocity dispersion ($\sigma$). We note that in order to feed our reconstructed cube to {$^{\rm 3D}${\sc Barolo}} we first need to interpolate it onto a regular grid of pixels.

In Figure~\ref{fig:dynamical_modelling_figure_1} we show the results of our dynamical modelling (see Appendix~\ref{sec:Appendix_B} for details of the modelling process). Overall, the model reproduces the observed velocity map (top panels in Figure~\ref{fig:dynamical_modelling_figure_1}) but we notice some residual velocity components ($\sim$ 50 $\rm km \, s^{-1}$) which are co-spatial with the leading sides of the bar (we show in each panel contours of the stellar continuum emission, which are the same as in Figure~\ref{fig:source_plane_reconstructions}). We interpret these residuals as evidence for non-circular motions, as expected for motions due to the presence of a bar. In the middle and bottom rows of Figure~\ref{fig:dynamical_modelling_figure_1} we show position-velocity (PV) diagrams along the kinematic major and minor axes, respectively. The PV diagram along the kinematic major axis displays the characteristic S-shape indicative of a rotating disc. The PV diagram along the minor axis has a diamond shape, again a feature seen in rotating disc galaxies.

Inspecting the best-fitting parameters of each of the fitted rings we find a mild upwards trend with increasing radius for the inclinations, varying between 48 and 50 degrees. On the other hand, the position angle remains almost constant at $\sim$119 degrees. 
The inclination-corrected rotational velocity increases out to a radius of $r \sim 1.8$~kpc and then flattens. We use the flat outer part of the rotation curve, specifically the five outer rings, to estimate an average value for the maximum rotational velocity of, $V_{\rm max} = 313 \pm 5 \, {\rm km \, s^{-1}}$. The best-fit velocity dispersions of the outer rings appear to continue to decrease, where the average value of the 5 outer rings is, $\sigma = 32 \pm 4 \, {\rm km \, s^{-1}}$. From these two values above, we estimate a ratio of rotational velocity to velocity dispersion of, $V / \sigma = 9.8 \pm 1.2$. We discuss in Section~\ref{sec:discussion_dynamics} how the dynamical properties of our source compare with the properties of its population as well as more typical star-forming galaxies at high redshift.

Finally, we estimate the dynamical mass of the source, ${M}_{\rm dyn}$, which is a measure of the total mass (baryonic + dark) within a given radius, $r$, using the formula,
\begin{equation}
    {M}_{\rm dyn} = 2.32 \times 10^5 \left(\frac{{V}_{\rm c}}{\rm km \, s^{-1}}\right)^2 \left(\frac{r}{\rm kpc}\right) \,\, {\rm M}_{\odot},
\end{equation}
where $V_{\rm c}$ is the circular velocity and $r$ is the radius at which we compute the dynamical mass. The circular velocity was computed directly in {$^{\rm 3D}${\sc Barolo}} by enabling the asymmetric drift correction functionality \citep[i.e. correction for pressure support][]{2007ApJ...657..773V}, and the average value of the 5 outer rings is, ${V}_{\rm c} = 327 \pm 3 \, {\rm km \, s^{-1}}$. Our estimate of the dynamical mass out to a radius, $r = 4$ kpc\footnote{Almost $100$~per~cent of the F444W flux is within $r < 4$~kpc (see Figure~\ref{fig:COG}), and so we assume that all of the stellar mass is contained within this radius.}, which is the extent of the region where we probe the dynamics of our galaxy (see Figure~\ref{fig:source_plane_mom_1_reconstructions}), is ${M}_{\rm dyn} = \left( 9.7 \pm 2.0 \right) \times 10^{10}\,{\rm M}_{\odot}$. This value is smaller than the value reported by \cite{2022A&A...663A..22G}, ${M}_{\rm dyn} = \left( 25 \pm 4 \right) \times 10^{10}\,{\rm M}_{\odot}$, who, however, used lower resolution CO(7-6) data and did not perform the full 3D modelling analysis to infer the inclination directly from the kinematics. Instead, they used the axis ratio as an estimate for the inclination, quoting a value of $\sim39$ deg, which is the origin of the discrepancy between the two estimates.

\subsubsection{The dynamics of SPT\protect\nobreakdash-2147}\label{sec:discussion_dynamics}

The dynamics of massive dusty star-forming galaxies at high redshift has been the focus of many studies dating back several years \citep[e.g.][]{2012ApJ...760...11H, 2014A&A...565A..59D} as well as more recently \citep[e.g.][]{2018ApJ...863...56C, 2020Natur.584..201R, 2021MNRAS.507.3952R, 2021A&A...647A.194F, 2021Sci...371..713L, 2021MNRAS.503.5329H, 2023MNRAS.521.1045R} exploiting data of higher resolution and sensitivity. One of the main questions that these studies have addressed is whether the dynamics of DSFGs are consistent with a rotating disc or an interaction/merger. The answer may also shed light on the mechanism that supports their high star formation rates. Many previous studies relied on low resolution observations, which can complicate the interpretation due to beam smearing effects \citep[e.g.][]{2022A&A...667A...5R}, which may cause mergers look like rotating disc galaxies. Now, however, the number of sources with data of sufficient quality for such analyses is large enough that we are beginning to see that both of these mechanisms are at play, i.e. rotating discs \citep[e.g.][]{2021MNRAS.507.3952R} and interacting/merging galaxies \citep[e.g.][]{2020MNRAS.494.5542R}, where for the latter case we might also see the companion galaxy at other wavelengths \citep[e.g. SPT-0418;][]{2023ApJ...944L..36P, 2023arXiv230710115C}. SPT\nobreakdash-2147 falls in the first category, based on its reconstructed velocity and velocity dispersion maps (Figure~\ref{fig:source_plane_mom_1_reconstructions}), which are well resolved given that $\sim$20 independent resolution elements can fit across the area of the source. While we can confidently state that the source did not experience a close-proximity interaction with another massive galaxy in its recent past (which could have disrupted its ordered circular motions, provided the companion's mass was significant enough), we cannot rule out the possibility of a scenario involving a more distant and less massive companion galaxy interacting with SPT\nobreakdash-2147 (see Section~\ref{sec:discussion_spiral}). This is because the magnification decreases as we move further away from the caustic curve, implying that we would likely not detect such a companion further away.

When the source is involved in a merger, often causing irregularities in its velocity map, extracting meaningful information becomes challenging since mergers are typically chaotic and difficult to model. However, in the case of a rotating disc galaxy, numerous properties can be derived by modeling its dynamics (Section~\ref{sec:dyn_mod}), offering valuable insights into the physical processes at play. A surprising recent discovery within this context is that DSFGs, often perceived as hosting violent conditions due to their intense star formation, apparently possess thin and cold gas discs \citep[e.g.,][]{2021MNRAS.507.3952R, 2021Sci...371..713L}, exhibiting remarkably low velocity dispersions, $\sigma  = 10 - 40 \, {\rm km \, s^{-1}}$. This contradicts previous observational studies of more `typical' star-forming galaxies that suggest a median velocity dispersion of $\sim40 \, {\rm km \, s^{-1}}$ at $z \sim 2.3$ \citep[e.g.][]{2017ApJ...842..121U}, but with individual measurements often reaching values of $\sim 100 \, {\rm km \, s^{-1}}$.

For SPT\nobreakdash-2147 we estimate a velocity dispersion, from [C{\sc ii}], of $\sigma = 32 \pm 4 \, {\rm km \, s^{-1}}$, in line with previous estimates for the SMG/DSFG population. As mentioned before, this measurement results in a value for the ratio of rotational velocity to velocity dispersion of, $V / \sigma = 9.8 \pm 1.2$. Based on this estimate, our source is classified as being rotation dominated, again in line with previous such estimates for this population, $V/\sigma = 8 - 10$ \citep[e.g.][]{2021MNRAS.507.3952R, 2021Sci...371..713L}. One possible way to reconcile the measured velocity dispersions for DSFGs and more typical SFGs is the tracer used to measure this quantity, [C{\sc ii}] line emission versus H$\alpha$. It is possible that the H$\alpha$ emission is associated with a thicker, hotter gaseous disc that is also more turbulent, while the [C{\sc ii}] emission line traces a colder thin disc. There are not many examples for which we have measurements of both these lines for galaxies at high redshift \citep[e.g.][]{2018ApJ...854L..24U}, at comparable spatial and spectral resolution, and so this interpretation still awaits confirmation. Unfortunately, we can not carry out this experiment due to the low velocity resolution of NIRSpec, $\sim 100 \, {\rm km \, s^{-1}}$. A discussion of the mechanism responsible for such cold discs in systems with high star formation activity is outside the scope of this work, but we refer the interested reader to \cite{2021MNRAS.507.3952R} and references therein.

Correcting our measured stellar (from SED; see Section~\ref{sec:SED}) and gas masses for magnification we estimate a total baryonic mass (stars $+$ gas) of ${M}_{\rm b} = \left( 9.1 \pm 1.1 \right) \times 10^{10}$~${\rm M}_{\odot}$. Using the dynamical mass as a measure of the total mass of the galaxy we estimate a ratio of baryonic to total mass, $M_{\rm bar} / M_{\rm dyn} = 0.9 \pm 0.2$, which is consistent with our source being baryon dominated out to a radius of $r = 4$~${\rm kpc}$. In order to make this estimate we assume that the stellar and gas masses that we measure for our source, which are based on integrated galaxy properties (i.e. observed total fluxes), correspond to their total values within the aperture used to measure the dynamical mass, 4.0~kpc. This is not necessarily true as the molecular gas distribution could extend beyond 4.0~kpc, since it is based on unresolved observations of the CO(2-1) emission line. Therefore, the ratio of baryonic to total mass should be considered as an upper limit. Taking into account the uncertainties in this measurement, the dark matter fraction, $f_{\rm DM}$, within 4.0~kpc is $<30$~per~cent.

\section{Discussion} \label{sec:section_5}

In this section, we discuss the morphological characteristics (i.e. bars and spiral arms) of SPT\nobreakdash-2147 (see previous sections) and place these within the wider context of high-redshift SMG/DSFG galaxy populations. We also discuss whether the presence of these morphological features are predicted in theoretical studies of galaxy formation and evolution.

\subsection{Bars}\label{sec:discussion_bar}

\begin{figure*}
    \centering
    \begin{tabular}{cc}
    \includegraphics[width=0.3\textwidth,height=0.325\textheight]{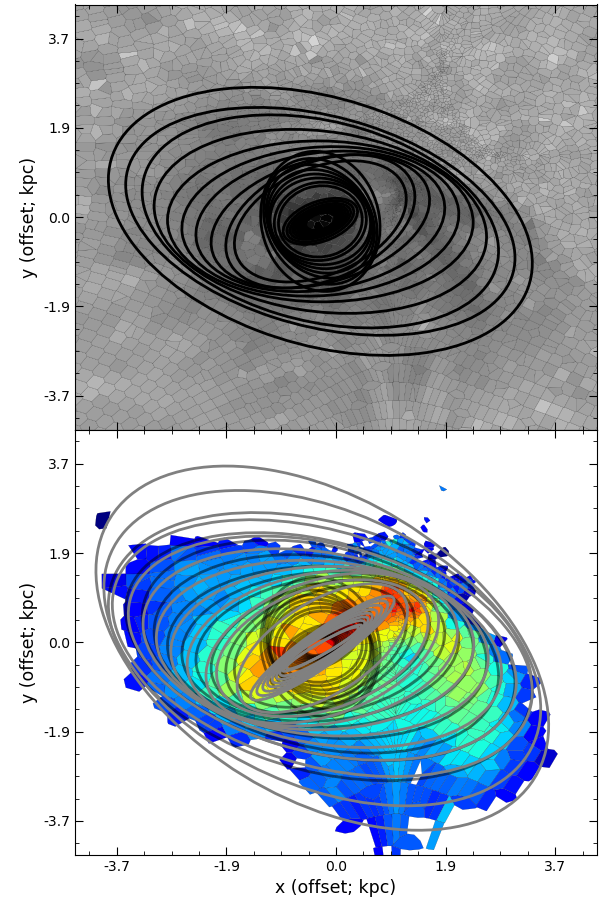} & \includegraphics[width=0.6\textwidth,height=0.325\textheight]{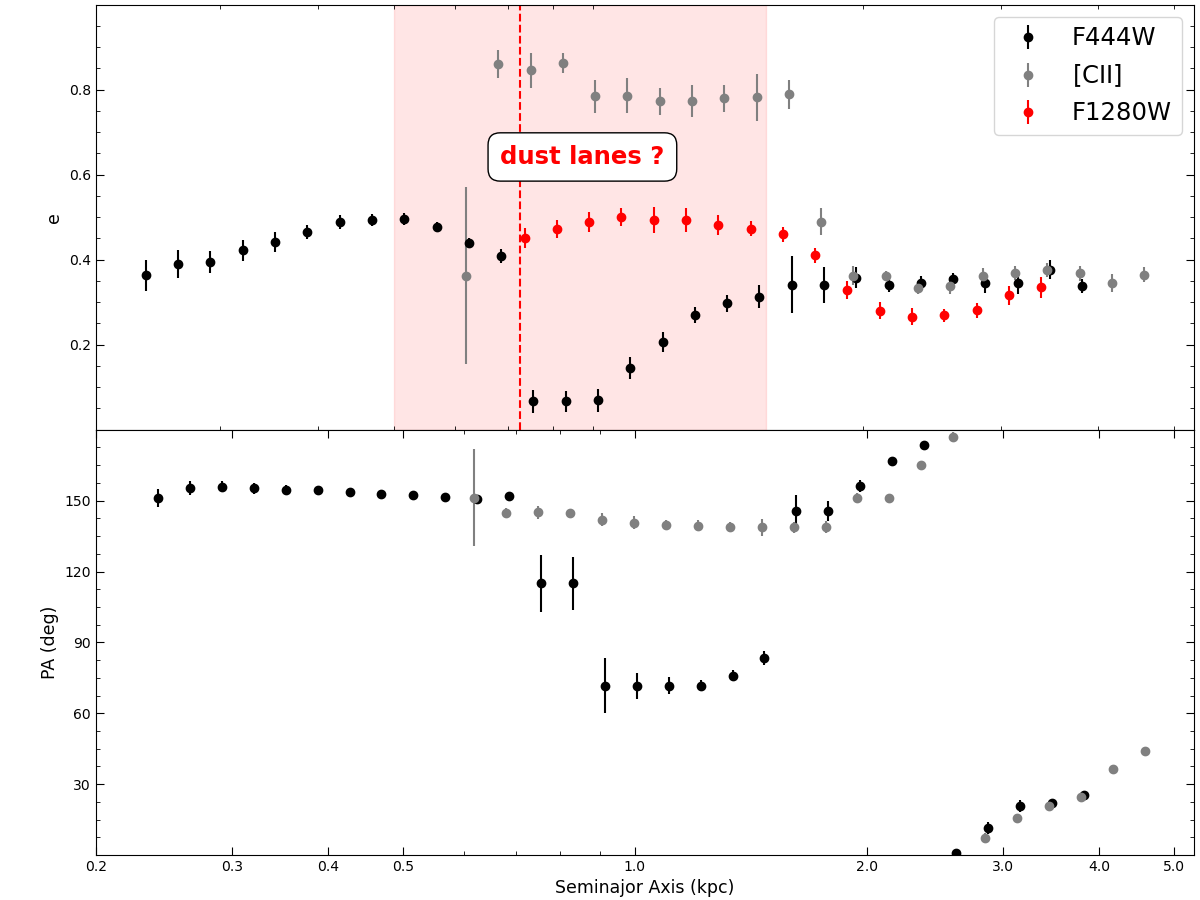} \\
    \end{tabular}
    
    \caption{Ellipse fitting to the reconstructed F444W/F1280W stellar continuum and [C{\sc ii}] emission line distributions. The top and bottom left panels shows the F444W and [C{\sc ii}] reconstruction, which are the same as in Figure~\ref{fig:source_plane_reconstructions}, with the best-fitting ellipses superimposed. The right panels show the radial profiles for the ellipticity (e) and position angle (PA) plotted against the semimajor axis (derived from the ellipse fitting procedure), in the top and bottom panels, respectively, for the F444W (black), F1280W (red) and [C{\sc ii}] (gray). The vertical red dashed line in the top panel indicates the lower limit set by the resolution in the F1280W filter. The outer ellipses, describing the extended disc ($r > 2$~kpc), agree perfectly between all tracers, however, the ellipses fitted to the central region show some differences between stars and gas, which could be attributed to dust attenuation (see main text).
    }
    \label{fig:ellipse_fitting}
\end{figure*}

Bars are ubiquitous in local spiral galaxies, with about two thirds of them having bar-like structures at their centre \citep[e.g][]{1993RPPh...56..173S, 2000AJ....119..536E}. While bars in galaxies are typically identified through the distribution of stars, many galaxies also exhibit a bar structure in their gas distribution \citep[e.g.][]{2021ApJS..257...43L, 2023arXiv230517172S}. These bars are believed to play a crucial role in the evolution of galaxies by facilitating more efficient gas flow towards the center through gravitational torques and shocks \citep[e.g.][]{2005MNRAS.358.1477A}. Despite their prevalence in local galaxies, the exact formation time of these structures remains uncertain.

Searching for bars in higher redshift galaxies has so far been challenging, mainly due to observational limitations. Before the launch of JWST most such studies relied on observations with HST, which are limited to rest-frame optical wavelengths for redshifts $z > 1$. In studies of local galaxies the fraction of identifiable bars in galaxies increases when analyzing images taken in the rest-frame NIR compared to optical wavelengths \citep[e.g.][]{2022ApJ...937L..33S}. This is because observations at these wavelengths are better tracers of the underlying stellar distributions since they are less affected by dust extinction \citep[][Knapen et al. 2000]{2007ApJ...657..790M}. It is therefore not surprising that we have only been able to detect bars in galaxies out to $z \sim$ 1 when using HST observations \citep[e.g.][]{2004ApJ...615L.105J, 2014MNRAS.438.2882M, 2014MNRAS.445.3466S}. Since the launch of JWST, access to rest-frame NIR wavelengths out to higher redshifts make it possible to identify bars in galaxies. For example, \cite{2023ApJ...945L..10G} have discovered good evidence of bars out to $z \sim 2.3$ using JWST observations obtained as part of the Cosmic Evolution Early Release Science Survey \citep[CEERS; see also][]{2023arXiv230616039S, 2023arXiv230910038L, 2023Natur.623..499C}.

Our reconstructions of the stellar continuum in the F444W filter and the [C{\sc ii}] emission line distribution (see Figure~\ref{fig:source_plane_reconstructions}) show evidence of a linear structure at the centre of our source. The source reconstructions using the MIRI bands also appear elongated along the same direction, but due to resolution limitations the feature is not fully resolved. We find that the linear structure is distinct from the extended disc and it is almost aligned with the kinematic minor axis (see Figure~\ref{fig:source_plane_mom_1_reconstructions}). It also seems that spiral arms originate from the ends of the linear feature, as is typically seen in barred spiral galaxies in the local Universe. Although typically bars are identified in galaxies by inspecting the distribution of stars, as mentioned before, there are numerous examples in the literature of well-defined bars in the gas distribution. Given the high gas fraction that we measure for SPT\nobreakdash-2147, it is not surprising that the bar is more clearly seen in the gas distribution.

Similarly to \cite{2023ApJ...945L..10G}, we apply an ellipse fitting technique to the reconstructed F444W, F1280W and [C{\sc ii}] emission line distributions. The goal is to determine the radial variation of the ellipticity and position angle of the ellipses and assess whether they display similar characteristics to sources that unambiguously show a bar in the stellar distributions. In addition, this ellipse fitting technique allows us to put a constraint on the length of the bar.

We make use of the \text{``isophote.Ellipse.fit{\_}image''} in {\sc PHOTUTILS} from {\sc Astropy} package (Bradley et al. 2020). We apply it to our reconstructions after interpolating them onto a regular grid. We first run the ellipse fitting on the reconstructed F444W distribution allowing all parameters, including the centre, to be free. We then run the algorithm a second time, fixing the centre of the ellipses to the median value estimated in the initial run. Having determined the centre of the ellipses, we then run the algorithm on the [C{\sc ii}] reconstruction again fixing the centre to be the same as those of the F444W reconstruction. Finally, for the ellipse fitting to the MIRI F1280W reconstruction, we follow the same two-step procedure that we did for the F444W reconstruction. The results of our ellipse fitting procedure are shown in Figure~\ref{fig:ellipse_fitting}.

\cite{2023ApJ...945L..10G}, as well as previous studies in the literature \citep[e.g.][]{2004ApJ...615L.105J, 2005MNRAS.364..283E, 2008MNRAS.384..420G}, considered a galaxy to be barred if it satisfies two criteria: (i) the ellipticity gradually increases in the bar-dominated region, reaching a maximum value $e_{\rm max} > 0.25$, while the position angle remains relatively constant. (ii) The ellipticity drops by at least 0.1 at the transition from the bar to the outer disc, while the position angle changes by at least 10~deg.
 
The ellipticity of the ellipses increases along the semi-major axis in the central region but there is a strong drop in ellipticity at 0.5 kpc and the orientation of the ellipses changes by $\sim 90$ deg. This behavior of the fitted ellipses to the F444W reconstruction is a consequence of a decrease in surface brightness on either side of the central region along the bar axis. {We speculate that this is a consequence of dust lanes running along the major-axis of the bar that obscure the stellar light emission. This is a common phenomenon in local barred galaxies, often seen in the leading edges of the bar}. The reason for attributing the decrease in surface brightness to dust is that the F1280W reconstruction does not show the same behavior, and in theory should be less affected by dust. In fact, fitting ellipses to the F1280W reconstruction reveals that the trend in ellipticity that is seen for the F444W ellipses continues inside this region (red points in the top panel of Figure~\ref{fig:ellipse_fitting}). We highlight (right upper panel in Figure~\ref{fig:ellipse_fitting}) the region where dust is obscuring the stellar light distribution. Finally, in the same region, the fitted ellipses to the [C{\sc ii}] emission line distribution appear to be very elliptical, $e \sim 0.8$, and this trend continues out to $r \sim 1.5$ kpc after which point the ellipticity drops by $\sim 0.3$. We consider this as evidence that the bar is seen in the gas distribution, perhaps even more clearly than in the stellar distribution. Considering all of the above, we estimate a bar length of $L_{\rm bar} \sim 1.5$~kpc, at which the ellipticity drops by more than 0.1 in the fits to both the [C{\sc ii}] and F1280W reconstructions.

If our interpretation of a bar in SPT\nobreakdash-2147 is correct, the question of when bar formation begins in galaxies arises naturally. Early works from \cite{1973ApJ...186..467O}, for example, have established that baryon-dominated disk galaxies tend to become violently unstable, leading to rapid formation of bars, with mergers and interactions also playing a significant role in this formation process \citep[e.g.][]{2016ApJ...821...90A, 2018ApJ...857....6L}. Theoretical works suggest that bar formation is considered more efficient in cold discs \citep[e.g.][]{1986MNRAS.221..213A, 2013MNRAS.429.1949A}, however, this is not an absolute condition as recent studies showed that galaxies remain prone to bar formation even if the disc is dominated by a thick component \citep[][]{2023A&A...674A.128G}, as are the majority of discs in high-z galaxies. The exact formation timescales for bars have not been thoroughly explored under a variety of conditions (e.g., gas fractions, level of turbulence), but in some idealised simulations, trends have already been established, such as disc formation timescales versus disc mass fraction, $f_{\rm disc}$ \citep[][]{2018MNRAS.477.1451F, 2023ApJ...947...80B}.

To compare with these idealised studies, we constrain the disc mass fraction in SPT\nobreakdash-2147 in the range of, $f_{\rm disc} = 50 - 80$~per~cent, based on estimates of the stellar, gas and dynamical masses. This suggests that SPT\nobreakdash-2147 has enough time to form a bar prior to being observed at redshift, $z = 3.76$, as long as the disc formed $\sim 1$~Gyr before it was observed, which roughly agrees with our estimate of the age of the stellar population from our SED fitting analysis (see Section~\ref{sec:SED}), which is, however, a very uncertain quantity. We note that these idealised simulations only explored the formation of bars in galaxies with low gas fractions, $f_{\rm gas} < 0.1$, but for SPT\nobreakdash-2147 we measure a gas fraction of $f_{\rm gas} = 0.45 \pm 0.12$. { The formation of bars in galaxies with high gas fraction ($>$10\%) was only recently explored for the first time \citep[][]{2024arXiv240206060B}. In this work the authors find that high gas fraction in galaxies can accelerate the formation of bars, contrary to what was previously thought \citep[][]{2013MNRAS.429.1949A}, albeit the later work only consider galaxies with low gas fractions.} 

\subsection{Spiral-arms}\label{sec:discussion_spiral}
Spiral structures in late-type galaxies are common in the local Universe. However, it is still unclear when these structures, either in stars or gas, were initially formed. There have been several studies searching for spiral structures in galaxies at redshifts $z > 2$ in an effort to discover when these structures first formed, what their formation mechanism is \citep[e.g.][]{2018MNRAS.478..932H} and how they can potentially impact star formation \citep[][]{2012MNRAS.426..701M}. So far only a small number of galaxies have been identified at these redshifts \citep[][]{2003AJ....125.1236D, 2012Natur.487..338L, 2021Sci...372.1201T, 2022ApJ...939L...7C, 2022ApJ...938L..24F, 2023ApJ...942L...1W}, with the currently highest redshift corresponding to a submillimeter galaxy at $z \sim 4.4$ \citep[]{2021Sci...372.1201T}

Our lensing reconstructions of the distributions of stars and gas provide evidence for the presence of at least one spiral arm in SPT\nobreakdash-2147. As mentioned in Section~\ref{sec:morphologies}, we find a better-defined spiral structure east of the centre of the galaxy both in the stellar continuum as well as the H$\alpha$ emission line distributions. There is emission located northwest of the centre of the galaxy in the stellar distribution, where we would expect to see a secondary arm, but very little H$\alpha$ emission is associated with that structure, possibly due to the lower magnification in that region of the source plane. The previously highest redshift spiral galaxy detected in both stellar continuum and H$\alpha$ shows features in its distributions that are reminiscent of our reconstructions \citep[][]{2012Natur.487..338L}.

Assuming that the emission northwest of the centre of this galaxy is not a spiral arm but just part of the extended disc then this source would be classified as an $m=1$ spiral. This could be an indication that our source had a recent interaction with a companion galaxy that led to the formation of this structure. However, there is currently no sufficient evidence to support this scenario as we do not see a companion galaxy in any of our observations, and the velocity and velocity dispersion maps lack any irregularities that could support this hypothesis.

The presence of spiral arms in massive dusty star-forming galaxies was put forward as a possible interpretation of high-resolution ($\sim 0.5$ kpc) images of the dust continuum emission in non-lensed galaxies from the ALESS survey \citep[][]{2019ApJ...876..130H}. Even though our dust continuum reconstructions achieve a similar resolution in the source plane to that of observations presented by \cite{2019ApJ...876..130H}, we do not discern any such structures. However, we note that $\sim 20$~per~cent of the dust continuum flux is associated with a more extended component with an effective radius that is comparable to that of the extended stellar component. This implies that dust could be associated with the spiral arm structure but it is not bright enough for our current observations to be able to study its morphology.

\section{Conclusions} \label{sec:section_6}

In this work, we have combined datasets across a wide range of the electromagnetic spectrum to present the most comprehensive view of a galaxy at redshift $z \sim 3.7$ utilizing the magnifying power that strong gravitational lensing offers. Our datasets, which come from a variety of instruments (newly obtained from JWST and archival from HST and ALMA), trace different components of the galaxy, including stars, dust and gas (ionized/molecular/atomic). In addition, observations of the [C{\sc ii}] emission line allow us to study the kinematics of this galaxy and constrain its dynamical properties. Our main findings are summarized below:

\begin{itemize}
    \item Using all of the available photometry from JWST, as well as previously measured fluxes at far-infrared/submm wavelengths, we were able to place robust constraints on the stellar mass of SPT\nobreakdash-2147: $M_{\star} = (6.3 \pm 0.9) \times \, 10^{10}$~${\rm M}_{\odot}$ with standard assumptions for the IMF. In addition, using previously published measurements of the CO(2$-$1) line luminosity we measure a molecular gas mass of, $M_{\rm gas} = (2.8 \pm 0.7) \times \, 10^{10}$~${\rm M}_{\odot}$, again using standard assumptions.

    \item Using our reconstructions of the [C{\sc ii}] emission line we study the kinematics and show that SPT\nobreakdash-2147 is consistent with a rotating disc. In the outer part, the disc has a circular velocity (corrected for inclination and pressure support) of $V_{\rm c} = 327 \pm 3 \, {\rm km \, s^{-1}}$ and a velocity dispersion, $\sigma = 32 \pm 4 \, {\rm km \, s^{-1}}$. From the inferred dynamical properties of this source we estimate a dynamical mass of ${M}_{\rm dyn} = \left( 9.7 \pm 2.0 \right) \times 10^{10}$~${\rm M}_{\odot}$ within 4 kpc.
    
    \item We find evidence from the stellar and gas morphologies that supports a scenario where SPT\nobreakdash-2147 is a barred spiral galaxy, featuring at least one spiral arm originating from the edge of the bar. One spiral arm is clearly seen in H$\alpha$, east from the centre of the galaxy. The base of this spiral arm is also detected in the stellar light distribution (F444W). There is emission in the F444W reconstruction where we would expect to see the secondary spiral arms, but it is not high SNR perhaps due to the lower lensing amplification in this region of the source plane.

    \item The formation of bars so early in the history of the Universe (age$\,\sim\,$1.6 Gyr at $z = 3.76$) should not be a surprise based on previous studies. Theory have shown that baryon-dominated galaxies with a cold disc component, such as SPT\nobreakdash-2147 ($f_{\rm DM} < 30$~per~cent), quickly become unstable to bar formation (see discussion). Even cosmological simulations are now beginning to resolve such structures in galaxies as early as $z \sim 4$ \citep[e.g.][]{2022MNRAS.512.5339R}. We make a rough estimate for the bar formation timescale in SPT\nobreakdash-2147 considering the age of the stellar population we measured for our SED analysis ($\rm age \sim 0.8$ Gyr) as an indicator for when the disc formed. Assuming that bar formation can only occur subsequent to disc formation \citep[][]{2023ApJ...947...80B, 2024arXiv240206060B} this suggests a tentative timescale for the onset of bar formation of $t_{\rm form} \sim 0.8$~Gyr in SPT\nobreakdash-2147.
\end{itemize}
\section*{Acknowledgements}
AA, SC, CSF, QH and KO are supported by ERC Advanced Investigator grant, DMIDAS [GA 786910], to C.S.\ Frenk. KO acknowledges support by STFC through grant ST/T000244/1. This paper makes use of both the DiRAC Data-Centric system and the Cambridge Service for Data Driven Discovery (CSD3), project code dp004 and dp195, which are operated by the Institute for Computational Cosmology at Durham University and the University of Cambridge on behalf of the STFC DiRAC HPC Facility (www.dirac.ac.uk). These were funded by BIS National E-infrastructure capital grant ST/K00042X/1, STFC capital grants ST/H008519/1, ST/K00087X/1, ST/P002307/1, ST/R002425/1, STFC DiRAC Operations grant ST/K003267/1, and Durham University. DiRAC is part of the National E-Infrastructure.

This paper makes use of the following ALMA data: ADS/JAO.ALMA \# 2018.1.01060.S, 2019.1.00471.S ALMA is a partnership of ESO (representing its member states), NSF (USA) and NINS (Japan), together with NRC (Canada), MOST and ASIAA (Taiwan), and KASI (Republic of Korea), in cooperation with the Republic of Chile. The Joint ALMA Observatory is operated by ESO, AUI/NRAO and NAOJ.

\section*{DATA AVAILABILITY}
All data used in this work are publicly available.

\section*{Software Citations}

This work uses the following software packages:

\begin{itemize}

\item
\href{https://github.com/astropy/astropy}{{Astropy}}
\citep{astropy1, astropy2}

\item
\href{https://bitbucket.org/bdiemer/colossus/src/master/}{{Colossus}}
\citep{colossus}

\item
\href{https://github.com/dfm/corner.py}{{Corner.py}}
\citep{corner}

\item
\href{https://github.com/joshspeagle/dynesty}{{Dynesty}}
\citep{dynesty}

\item
\href{https://github.com/matplotlib/matplotlib}{{Matplotlib}}
\citep{matplotlib}

\item
\href{numba` https://github.com/numba/numba}{{Numba}}
\citep{numba}

\item
\href{https://github.com/numpy/numpy}{{NumPy}}
\citep{numpy}

\item
\href{https://github.com/rhayes777/PyAutoFit}{{PyAutoFit}}
\citep{pyautofit}

\item
\href{https://github.com/Jammy2211/PyAutoGalaxy}{{PyAutoGalaxy}}
\citep{pyautogalaxy}

\item
\href{https://github.com/Jammy2211/PyAutoLens}{{PyAutoLens}}
\citep{2018MNRAS.478.4738N, 2021JOSS....6.2825N}

\item
\href{https://github.com/jyhmiinlin/pynufft}{PyNUFFT}
\citep{pynufft}

\item
\href{https://www.python.org/}{{Python}}
\citep{python}

\item
\href{https://github.com/scikit-image/scikit-image}{{Scikit-image}}
\citep{scikit-image}

\item
\href{https://github.com/scikit-learn/scikit-learn}{{Scikit-learn}}
\citep{scikit-learn}

\item
\href{https://github.com/scipy/scipy}{{Scipy}}
\citep{scipy}

\end{itemize}


\bibliographystyle{mnras}
\bibliography{main_SPT2147}

\appendix

\section{Lensed source emission in HST image} \label{sec:Appendix_A}

\begin{figure*}
    \centering
    \includegraphics[width=0.95\textwidth,height=0.185\textheight]{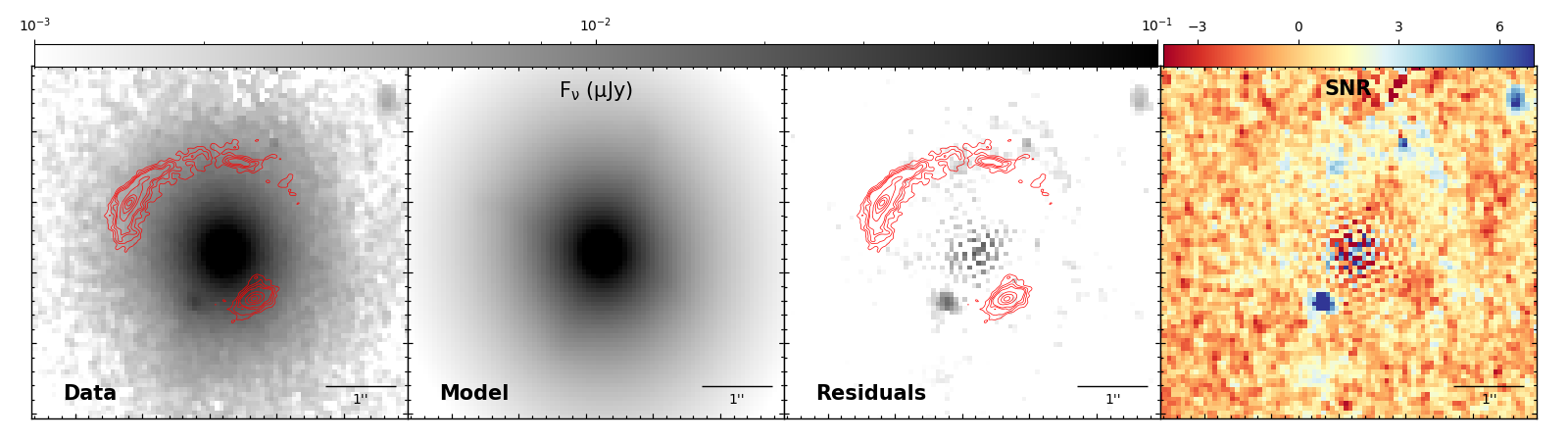} 
    \caption{Results from the lens light subtraction analysis of the HST F140W image for SPT\nobreakdash-2147. The panels from left to right show the observed image, best-fitting model, residuals and the SNR map (ratio of the residual image and the noise map). The red contours on the first and third panels of this figure correspond to the ALMA band 8 dust continuum emission and are the same as the bottom left panel in Figure~\ref{fig:fig1}.}
    \label{fig:hst_lens_light_subtraction_subplots}
\end{figure*}

We attempt to reveal the potential presence of emission from the background source in the HST image by fitting the lens's light using a double S\'{e}rsic profile (see method in Section~\ref{sec:section_3}) and then subtracting it from the observed data. The results from our lens light subtraction analysis are shown in Figure~\ref{fig:hst_lens_light_subtraction_subplots}, where the different panels from left to right are the observed image, the best-fitting model of the lens's light emission, residuals and the signal-to-noise ratio (SNR) map of the residual emission. Although the emission from the background source seen by HST is not used in our analysis, due to its low SNR, this goes to demonstrate the need for observations in the rest-frame NIR.

\section{Details of kinematic modelling} \label{sec:Appendix_B}

We give the detailed configuration used for {$^{\rm \sc 3D}${\sc Barolo}} in Table~\ref{tab:3db_config}. In addition to this fiducial configuration used throughout our analysis above, we also ran several variant models to check for sensitivity of our conclusions to choices made in the modelling process. In most cases the influence of alternative assumptions was small, e.g. $\lesssim 10$~per~cent in the rotation curve amplitude at fixed radius. These variations included changing the figure of merit in the optimization process (parameter {\sc ftype}), the flux normalization scheme ({\sc norm}), and the functional form of the geometric regularization ({\sc regtype}).

Two variations, however, led to larger changes in the model rotation curve. By default, {$^{\rm \sc 3D}${\sc Barolo}} weights pixels according to their angular distance $\theta_{\rm maj}$ from the kinematic major axis as $|\cos\theta_{\rm maj}|$. Weighting more strongly ($\cos^2\theta_{\rm maj}$) makes little difference, but using a uniform weighting for all pixels leads to a preference for a lower inclination and a higher rotation curve amplitude, with a peak rotation velocity of nearly $400\,\mathrm{km}\,\mathrm{s}^{-1}$ (with the change in rotation speed largely due to the change in the inclination). We suspect that the models weighted by $\theta_{\rm maj}$ better capture the circular velocity curve of the galaxy, since the unweighted model is likely biased by the bar (roughly aligned with the minor axis) -- residuals due to the bar are clearly visible along the kinematic minor axis (Fig.~\ref{fig:dynamical_modelling_figure_1}, right panel). The degeneracy between a bar pattern and the inclination angle (and consequently the rotation speed) is well-known \citep[e.g.][]{1997MNRAS.292..349S}.

We also experimented with allowing a radial flow component in the model (parameter {\sc vrad}), but found that the outcome was very sensitive to the initial guess provided. Given the strong degeneracies between inclination angle, a bar pattern and radial flows, it is unsurprising that leaving all of these as free parameters causes the model to lose constraining power. Overall, our impression is that the amplitude of the flat portion of the rotation curve in our fiducial model -- our main interest in this work -- is likely robustly estimated by our modelling process, but that our model (locally axisymmetric) model fails to capture the detailed kinematics of the galaxy that are dominated by an $m=2$ non-axisymmetric feature. A model that explicitly includes such features \citep[e.g.][]{2015arXiv150907120S} may be more appropriate for further study of the kinematics of SPT\nobreakdash-2147.

\begin{table*}
\caption{Configuration used for {$^{\rm \sc 3D}${\sc Barolo}}, omitting those with no influence on the calculation (e.g. file paths, etc.). Other parameters omitted from this list adopt the default values set by {$^{\rm \sc 3D}${\sc Barolo}}.}
\label{tab:3db_config}
\begin{tabular}{lrlp{1.2\columnwidth}}
Parameter & Value & Units & Description \\
\hline
{\sc 3dfit} & {\sc true} & -- & Enable 3D fitting. \\
{\sc nradii} & {\sc 25} & -- & Number of rings. \\
{\sc radsep} & 0.025 & $\mathrm{arcsec}$ & Inter-ring spacing. \\
{\sc vsys} & 0.0 & $\mathrm{km}\,\mathrm{s}^{-1}$ & Initial guess for systemic velocity. \\
{\sc vrot} & 330.0 & $\mathrm{km}\,\mathrm{s}^{-1}$ & Initial guess for rotation speed. \\
{\sc vdisp} & 37.0 & $\mathrm{km}\,\mathrm{s}^{-1}$ & Initial guess for velocity dispersion. \\
{\sc vrad} & 0.0 & $\mathrm{km}\,\mathrm{s}^{-1}$ & Radial flows are not modelled. \\
{\sc inc} & 47.0 & $\mathrm{deg}$ & Initial guess for inclination. \\
{\sc pa} & 115.0 & $\mathrm{deg}$ & Initial guess for position angle. \\
{\sc z0} & 0.0138 & $\mathrm{arcsec}$ & Disc scale height, equivalent to $100\,\mathrm{pc}$ \citep[see][sec.~3.2.1.ii \& 7.1]{2017MNRAS.466.4159I}. \\
{\sc deltavsys} & 100.0 & $\mathrm{km}\,\mathrm{s}^{-1}$ & Allowed variation from initial guess for systemic velocity. \\
{\sc deltainc} & 15.0 & $\mathrm{deg}$ & Allowed variation from initial guess for inclination. \\
{\sc deltapa} & 30.0 & $\mathrm{deg}$ & Allowed variation from initial guess for position angle. \\
{\sc free} & {\sc vrot vdisp inc pa vsys} & -- & Free parameters of each ring: rotation speed, velocity dispersion, inclination, position angle, systemic velocity. Systemic velocity is fixed to a common constant for all rings to produce the final model.\\
{\sc ftype} & 2 & -- & Figure of merit for minimization set to $|{\rm model}-{\rm data}|$ \\
{\sc wfunc} & 2 & -- & Pixels weighted by angular distance $\theta_{\rm maj}$ from major axis as $\cos^2\theta_{\rm maj}$. \\
{\sc ltype} & 1 & -- & Disc has Gaussian vertical structure. \\
{\sc norm} & {\sc local} & -- & Flux in each pixel is normalised before fitting kinematic model. \\
{\sc twostage} & {\sc true} & -- & After initial fit, geometric parameters (systemic velocity, inclination, position angle) are regularised and remaining parameters are fit again. \\
{\sc regtype} & 1 & -- & Inclination and position angle regularised with a polynominal of degree 1. \\
{\sc bweight} & 1 & -- & Relative weight for pixels with zero flux. \\
{\sc flagerrors} & {\sc true} & -- & Statistical uncertainties are estimated \citep[see][sec.~2.5.v]{2015MNRAS.451.3021D}. \\
{\sc linear} & 0.4246 & $\mathrm{channels}$ & Equivalent instrumental root-mean-square spectral broadening. \\
{\sc adrift} & {\sc true} & -- & `Asymmetric drift correction' \citep[but see][appendix~A]{2007ApJ...657..773V} calculated. \\

\hline
\end{tabular}
\end{table*}


\end{document}